\documentclass[11pt]{article}
\usepackage[centertags]{amsmath}
\usepackage{amsfonts} \usepackage{amssymb} \usepackage{amsthm}
\usepackage{epsfig}\usepackage{stmaryrd}\usepackage{latexsym}
\newcommand{\bra}[1]{\langle{#1}|}
\newcommand{\ket}[1]{|{#1}\rangle}
\def\one{{\rm 1\kern -.9mm l}}                             %

\def\beq{\begin{equation}}
\def\eeq{\end{equation}}
\def\beqa{\begin{eqnarray}}
\def\eeqa{\end{eqnarray}}

\def\sgh{{\rm sgh}}

\def\R{{\rm R}}
\newcommand{\bea}{\begin{eqnarray}}
\newcommand{\ena}{\end{eqnarray}}

\usepackage{graphicx}
\usepackage{dcolumn}
\begin{document}

\begin{titlepage}
\rightline{DSF-13/2003}
\rightline{NORDITA-2003-23 HE} \vskip 3.0cm
\centerline{\LARGE \bf Gauge/Gravity Correspondence }\vskip .5cm
\centerline{\LARGE \bf from} \vskip .5cm
\centerline{\LARGE \bf Open/Closed String Duality  }
\vskip 1.4cm \centerline{\bf P. Di Vecchia
$^a$, A. Liccardo $^b$, R. Marotta $^b$ and F. Pezzella $^b$}
\vskip .8cm \centerline{\sl $^a$ NORDITA, Blegdamsvej 17, DK-2100 Copenhagen \O, Denmark}
\vskip .4cm \centerline{\sl
 $^b$ Dipartimento di
Scienze Fisiche, Universit\`a di Napoli and INFN, Sezione di
Napoli} \centerline{\sl Complesso Universitario Monte
S. Angelo, ed. G  - via Cintia -  I-80126 Napoli, Italy}
 \vskip
2cm

\begin{abstract}
We compute the annulus diagram corresponding to the interaction of a
fractional D3 brane with a gauge field on its world-volume  and a stack
of N fractional D3 branes on the orbifolds $\mathbb{C}^2 /\mathbb{Z}_2$ and
$\mathbb{C}^3/(\mathbb{Z}_2 \times \mathbb{Z}_2)$.
We show that its logarithmic divergence can
be equivalently understood as due either to massless open string
states circulating in the loop or to massless closed string states
exchanged between two boundary states. This follows from the fact
that, under open/closed string duality,
massless states in the open and closed string channels are matched
into each other without mixing with massive states.
This explains why
the perturbative properties of many gauge theories living on the
worldvolume of less supersymmetric and nonconformal branes have been recently
obtained from their corresponding supergravity solution.

\end{abstract}
\vfill  {\small Work partially supported by the European
Commission RTN Programme HPRN-CT-2000-00131 and by MIUR.}
\end{titlepage}

\newpage

\tableofcontents       %
\vskip 1cm             %

\section{Introduction}
\label{sez0}

A D brane has the twofold property of being a solution
of the low-energy string effective action, which is
just given by supergravity, and of having open
strings with their endpoints attached to its world-volume. In
particular, the lightest open string excitations correspond to a
gauge field and its supersymmetric partners if the theory is
supersymmetric. These complementary descriptions open the
way to study quantum properties of the gauge theory living in the
world-volume of a D brane from the classical dynamics of the brane and
viceversa. This goes under the name of
gauge/gravity correspondence that has allowed to derive properties
of ${\cal{N}}=4$ super Yang-Mills - as one can see for example in
Ref.~\cite{CJ} - and also, by the addition of a decoupling limit, to
formulate the Maldacena conjecture of the equivalence between
${\cal{N}}=4$ super Yang-Mills and type IIB string theory
compactified on $AdS_5 \otimes S_5$~\cite{MALDA}.

Although it has not been possible to extend the Maldacena conjecture
to non-conformal and less supersymmetric gauge theories,
a lot of a priori unexpected information on these theories has been
obtained from the gauge/gravity correspondence\footnote{For general
reviews on various approaches see for instance
Refs.~[3 $\div$ 6].}. In particular, it has been shown
that the classical supergravity solutions corresponding to fractional
and wrapped D branes encode perturbative properties of non conformal
and less supersymmetric gauge theories living on their world-volume, such as
the chiral and scale anomaly~\cite{KOW,ANOMA}. It was of course
expected that those
properties could be derived by studying, in string theory, the gauge
theory living on the above D branes by taking the field theory limit of the
one-loop open string annulus diagram, with
such methods as those described for instance in Ref.~\cite{FILIM}. But it came
as a surprise that these  properties were also encoded in the
supergravity solution, especially after the formulation of the
Maldacena conjecture, which limited the validity of the supergravity
approximation  to the strong coupling regime of the gauge theory.

Being gauge theories and supergravity related to open and
closed strings respectively, which in turn are connected by the
open/closed string duality, there must be a relation between this latter and
the gauge/gravity correspondence.

In this paper we will use
fractional branes to analyze features of the relation between the
open/closed string duality and the gauge/gravity correspondence more deeply.
In particular, we will show that, working in the pure string framework
and using only the open/closed string duality,
the perturbative properties of the gauge theory living on a D brane
can be equivalently derived by performing the field theory limit
either in the open string channel, as expected, or in the closed string
channel, where the supergravity approximation holds.

We start by computing the one-loop open string annulus diagram which
describes the interaction of a fractional brane of the orbifold
$\mathbb{C}^2 /\mathbb{Z}_2$ having an
$SU(N)$
gauge field on it with $N$ fractional branes without any gauge field,
and we extract from it the coefficient of the gauge kinetic term. It
turns out, as  observed in Refs. [10 $\div$ 13],
that this contribution is logarithmically divergent already at the
string level. In general divergences higher than the logarithmic ones
correspond to gauge anomalies that make the theory inconsistent and
therefore  must be cancelled. Logarithmic divergences correspond instead
to the exchange of massless closed string states in the closed string
channel, as it can be seen from the fact that the boundary state has a 
non-vanishing coupling to them ({\em massless tadpoles}). Those divergences 
were originally found in the
bosonic string where they were caused by the dilaton exchange in the
closed string channel and were cured in different ways~[14
$\div$ 20]. In the case we are examining in
this paper they are due to the exchange, in the two directions transverse
to the branes and the orbifold, of massless twisted states
that have a nonzero coupling to  fractional D3 branes and therefore
contribute in the closed string channel. In general the presence of
tadpoles signal some kind of instability that must be cured.
On the other hand, it
turns out that those logarithmic divergences precisely correspond to
the one-loop divergences that one finds in the gauge theory living in
the world-volume of the brane. In this paper we just regularize the
string calculation
introducing an infrared cutoff in the closed string channel
corresponding to an ultraviolet cutoff in the open string channel and we show
that the divergent contribution can be seen to come either from the
exchange of massless closed string states between the two fractional branes
or from the massless open string states that go around the loop. Indeed,
under the modular transformation, which maps the
open to the closed string channel, {\em open string massless states go into
closed string massless states, and open string massive states go into
closed string massive states, without, surprisingly, showing any mixing
between massless and massive states}. By adding to the one-loop open
string diagram also the contribution of the open string tree diagrams, we
find an expression for the gauge coupling constant that gives the correct
beta-function of ${\cal N}=2$ SYM, exactly
reproducing what has been found from supergravity
calculations~[21 $\div$ 25].

Another important feature of our calculation is the appearance of
a non vanishing contribution from the Ramond
odd spin structure at the string level, giving the vacuum angle
$\theta_{YM}$ of the gauge theory living on the brane. The introduction of a
complex cutoff together with a symmetrization between the two
fractional branes occurring in the annulus diagram allows us
to reproduce the corresponding supergravity calculation.

The results obtained in the case of the orbifold $\mathbb{C}^2/\mathbb{Z}_2$
are then shown to hold also in the case of the orbifold
$\mathbb{C}^3/(\mathbb{Z}_2 \times \mathbb{Z}_2)$, where the world-volume gauge theory is ${\cal N}=1$ SYM.

In conclusion, in this paper we show
why supergravity calculations reproduce the
perturbative behavior of the gauge theory living on the fractional
branes of the orbifolds $\mathbb{C}^2 /\mathbb{Z}_2$ and
$\mathbb{C}^3/(\mathbb{Z}_2 \times \mathbb{Z}_2)$.
This is a consequence of the
fact that, when we introduce an external field, the annulus
diagram is divergent already at the
string level and of the open/closed string duality.

What remains still a bit obscure is why the logarithmic divergence that appears
{\em at string level} is directly related to the divergence
that one finds in the gauge theory living on the brane, reproducing the
correct field theory anomalies.

The paper is organized as follows. Section (\ref{sez1}) is devoted to
the calculation in the full string theory of the annulus diagram for
fractional branes of the orbifold $\mathbb{C}^2 /\mathbb{Z}_2$ as
a function of the gauge field on the dressed brane and of the distance
between the latter and the stack of the $N$ branes.
In Sect. (\ref{sez2}) the field theory limit is performed both in the
open and closed string channel and it is shown why the perturbative
properties of the gauge theory living on $N$ fractional D3 branes can
be derived from their supergravity solution. In Sect. (\ref{sez4}) we
extend the previous analysis to the orbifold
$\mathbb{C}^3 /(\mathbb{Z}_2 \times \mathbb{Z}_2) $
finding again agreement with the results obtained from the
supergravity solution for ${\cal{N}}=1$ supersymmetric gauge theories.
In Appendix \ref{app0} we list the $f$- and 
$\Theta$-functions and their
transformation properties under the modular
transformation. Appendix \ref{app2} is devoted to the calculation of the
annulus diagram in the open string channel, while that in the closed
string channel is performed in Appendix \ref{app4}.

\section{Branes in External Fields: ${\cal N}$ = 2 orbifold}
\label{sez1}

In this Section we analyze the interaction between a
D$3$ brane with an external $SU(N)$ gauge field on its world-volume (in the
following named {\em dressed} brane) and a stack of  $N$ ordinary
D$3$ branes. This can be equivalently
done either, in the open string channel, by computing the one-loop open string
diagram or, in the closed string channel, by computing the tree closed
string diagram containing two boundary states and a closed string
propagator.

In order to obtain a gauge theory with reduced supersymmetry we
consider type IIB superstring theory on an orbifold space.
Furthermore, to obtain a non conformal gauge theory, we study
fractional D3 branes\footnote{For a review of fractional branes
and their application to the study of non-conformal gauge theories
see for instance Ref.~\cite{REV}.} which are characterized by their
being stuck at the orbifold fixed point and which, unlike bulk branes,
have a non-conformal theory on their world-volume. For the sake of
simplicity we consider fractional branes of the orbifold ${\rm I\! R}^{1,5} \times
\mathbb{C}^2 /\mathbb{Z}_2$ that have ${\cal{N}}=2$ super Yang-Mills  on their
world-volume.

The $\mathbb{Z}_2$ group, that is chosen to be acting on the coordinates
$x^{m}$ with
$m=6,7,8,9$, is characterized by two elements $(e,h)$, being $e$
the identity element and $h$ such that $h^2=e$. The element
$h$ acts on the complex combinations
$\vec{z} = (z^1, z^2)$, where $z^1=x^6+ix^7$,
$z^2=x^8+ix^9$ as follows:
\begin{eqnarray}
(z_1 \,,\,z_2)\rightarrow ( - z_1 \,,\, - z_2).
\end{eqnarray}
The orbifold group $Z_2$ acts also on the Chan-Paton factors
located at the endpoints of the open string stretched between
the branes.  Fractional
branes are
defined as branes for which such factors transform according to irreducible
representations of the orbifold group and
we
consider only the trivial one corresponding to a particular kind of
fractional branes. The orbifold we are
considering is non compact and has therefore
only one fixed point located at $z_1=z_2=0$.
We are interested in the case of parallel fractional D3 branes with
their world-volume along the directions $x^0 , x^1 , x^2, x^3$, that are
completely external to the space on which the orbifold acts.
The gauge field lives on the four-dimensional world-volume of the
fractional D3 brane and can be chosen to have
the following form:
\begin{eqnarray}
\hat{F}_{\alpha \beta } \equiv
2 \pi \alpha' F_{\alpha \beta}  =  \left(
\begin{array}{cccc}
0  & f & 0 & 0\\
-f & 0 & 0 & 0\\
0  & 0 & 0 & g\\
0  & 0 &- g& 0
\end{array}
\right) \, .
\label{effe}
\end{eqnarray}

The interaction between two branes is given by the vacuum
fluctuation of an open string that is stretched between them.
In particular, the free energy of an open string between a D3 brane and a
stack of $N$ D3 branes located at a distance $y$ in the plane ($x^4 ,
x^5$) orthogonal to both the world-volume of the
 D3 branes and
the four-dimensional space on which the orbifold acts, is given by the
one-loop open string free energy:
\begin{eqnarray}
Z & = &N\int_{0}^{\infty} \frac{d \tau}{\tau} Tr_{{\rm NS-R}}
\left[ \left(\frac{e+h}{2}
\right) (-1)^{F_s}(-1)^{G_{bc}}\left( \frac{(-1)^{G_{\beta\gamma}} +
(-1)^F }{2} \right) { e}^{- 2 \pi \tau L_0} \right] \nonumber \\
& \equiv & Z_{e}^o + Z_{h}^o    \label{Z}
\end{eqnarray}
where $G_{bc}$ and $G_{\beta\gamma}$ are, respectively,
the ghost and superghost number operators, $F_s$ is
the space-time fermion number, $(-1)^{F}$ and $L_0$ are defined in Appendix B,
and the superscript $o$ stands for $open$.
The fact that we are considering a string theory on the
orbifold $\mathbb{C}^2 /\mathbb{Z}_2$ is encoded in the presence of the orbifold
projector
$P=(e+h)/2$ in the trace.
The explicit computation, shown in detail in Appendix~\ref{app2}, gives
the following results:
\begin{eqnarray}
Z_e^o & = & - \frac{N}{(8 \pi^2 \alpha')^{2}} \int d^4x
\sqrt{ -\mbox{det}(\eta+\hat{F})}
\int_{0}^\infty \frac{d\tau}{\tau}e^{-\frac{y^2 \tau}{2\pi\alpha'}}
\frac{\sin \pi \nu_f  \sin \pi \nu_g}{f_1^4(e^{-\pi \tau})
\Theta_{1}(i\nu_f\tau|i\tau)
  \Theta_{1}(i\nu_g\tau|i\tau)
} \nonumber \\
& & \times \left[f_3^4(e^{-\pi \tau})
\Theta_{3}(i\nu_f \tau|i \tau) \Theta_{3}(i\nu_g\tau|i\tau)
- f_4^4(e^{-\pi \tau})
\Theta_{4}(i\nu_f\tau|i\tau)\Theta_{4}(i\nu_g\tau|i\tau)  \right. \nonumber
\\
&&\left. - f_2^4(e^{-\pi \tau})
\Theta_{2}(i\nu_f \tau|i\tau)\Theta_{2}(i\nu_g\tau|i\tau) \right],
\label{zetae}
\end{eqnarray}

\begin{eqnarray}
Z_h^o & = & \frac{N}{(8 \pi^2 \alpha')^2} \int d^4x
\sqrt{ -\mbox{det}(\eta+\hat{F})}
\int_{0}^\infty \frac{d\tau}{\tau}e^{-\frac{y^2 \tau}{2\pi\alpha'}}
\frac{ 4\, \sin \pi \nu_f  \sin \pi \nu_g}
{\Theta_{2}^2(0|i\tau) \Theta_{1}(i\nu_f\tau|i\tau)\Theta_{1}
(i\nu_g\tau|i\tau)}
\nonumber \\
{} && \times \left[ \Theta_{3}^2(0|i\tau)\Theta_{4}(i\nu_f\tau|i\tau)
\Theta_{4}(i\nu_g\tau|i\tau) -
\Theta_{4}^2(0|i\tau) \Theta_{3}(i\nu_f\tau|i\tau)
\Theta_{3}(i\nu_g\tau|i\tau) \right] \nonumber \\
&&  + \frac{iN}{32\pi^2}\int d^4x\, F_{\alpha\beta}^a
{\tilde F}^{a\,\alpha\beta}
\int_{0}^\infty \frac{d\tau}{\tau}e^{-\frac{y^2 \tau}{2\pi\alpha'}},
\label{zetatet}
\end{eqnarray}
where $\tilde{F}_{\alpha \beta }=\frac{1}{2}\epsilon _{\alpha \beta \gamma
\delta }F^{\gamma \delta }$.
The $f$- and the $\Theta$-functions are listed in Appendix~\ref{app0}. In the
previous equations we have defined:
\begin{eqnarray}
\nu_{f} \equiv \frac{1}{2 \pi i}\log \frac{1+f}{1-f}
\,\,\,\,\,\,\,\,\,\,\,{\rm and}\,\,\,\,\,\,\,\,\,\,\,
\nu_{g} \equiv \frac{1}{2 \pi i}\log \frac{1-i g}{1+i g} .
\label{nufnug2}
\end{eqnarray}
The three terms in Eq. \eqref{zetae}  come  from the
 NS, NS$(-1)^F$, and R sectors respectively, while the contribution
from the R$(-1)^F$ sector vanishes. In Eq. \eqref{zetatet} the three terms
come from the NS$(-1)^F$, NS and R$(-1)^F$ sectors respectively, while
the $R$ contribution  vanishes because the projector $h$
annihilates the Ramond vacuum.

The above computation can also be  performed in the
{\em closed string channel} where $Z_e^c$ and $Z_h^c$ now
read as the tree level closed
string amplitude between two untwisted and two twisted boundary states
respectively, with the results:
\begin{eqnarray}
Z_e^c = \frac{ \alpha' \pi N }{2} \int_{0}^{+\infty} dt \, \, \, {}^U
\langle D3; F |
e^{-\pi t ( L_0 + \bar{L}_{0} - a_c) } |D3 \rangle^{U}
\label{ze32}
\end{eqnarray}
\begin{eqnarray}
Z_{h}^c = \frac{ \alpha'\pi N}{2} \int_{0}^{+\infty} dt \, \, \,
{}^T \langle D3; F | e^{- \pi t (L_0 + \bar{L}_{0}) } |D3
\rangle^{T} \label{zg32} \, ,
\end{eqnarray}
where $ |D3; F > $ is the boundary state dressed with the gauge field
$F$. The details of this calculation are presented in
Appendix~\ref{app4}. Here we give only the final results, i.e.:
\begin{eqnarray}
Z_e^c & = & \frac{N\, }{(8\pi^2\alpha')^2} \int d^4x
\sqrt{ -\mbox{det}(\eta+\hat{F})}
\int_{0}^\infty \frac{dt}{t^3}e^{-\frac{y^{2} }{2\pi\alpha't}}
\frac{\sin\pi\nu_f \sin\pi\nu_g}{
\Theta_{1}(\nu_f|it)\Theta_{1}(\nu_g|it)
f_1^4(e^{-\pi t})}
\nonumber\\
&& \times \left\{f_3^4(e^{-\pi t})
\Theta_{3}(\nu_f|it)\Theta_{3}(\nu_g|it)
-f_4^4(e^{-\pi t})\Theta_{4}(\nu_f|it)
\Theta_{4}(\nu_g |it) \right.\nonumber\\
&&\left.
- \Theta_{2}(\nu_f|it)\Theta_{2}(\nu_g|it)f_2^4(e^{-\pi t})\right\}~,
\label{op2}
\end{eqnarray}
\begin{eqnarray}
Z_h^c & = & \frac{N}{(8\pi^2\alpha')^2} \int d^4x
\sqrt{ -\mbox{det}(\eta+\hat{F})}
\int_{0}^\infty \frac{dt}{t}e^{-\frac{y^{2} }{2\pi\alpha't}}
\frac{4\,\sin\pi\nu_f \sin\pi\nu_g}{
\Theta_{4}^2(0|it) \Theta_{1}(\nu_f|it)\Theta_{1}(\nu_g|it)}
\nonumber\\
&&\times \left\{\Theta_{2}^2(0|it) \Theta_{3}(\nu_f|it)\Theta_{3}(\nu_g|it)
-\Theta_{3}^2(0|it) \Theta_{2}(\nu_f|it)\Theta_{2}(\nu_g|it) \right\}
\nonumber\\
&&+\frac{iN}{32\pi^2} \int d^4x
F^a_{\alpha\beta}\tilde F^{a\alpha\beta}\int \frac{dt}{t}
e^{-\frac{y^{2} }{2\pi\alpha't}}.
\label{closed}
\end{eqnarray}
The three terms in Eq. \eqref{op2} come from the NS-NS,
NS-NS$(-1)^ {F }$ and R-R sectors respectively, while those in
Eq.~(\ref{closed}) from the NS-NS, R-R and R-R$(-1)^{F }$
ones. In particular, the twisted odd  R-R$(-1)^{F}$ spin
structure gets a non vanishing contribution only from the zero
modes, as explicitly shown in Appendix~\ref{app4}.

It goes without saying that the two expressions for $Z$ separately obtained
in the open and  the closed string channels are as expected, equal
to each other. This equality goes under the name of
open/closed string duality and can be easily shown  by
using how  the $\Theta$ functions transform (see Eq. (\ref{modtras}))
under the modular transformation that relates the modular parameters
in the open and closed string channels, namely $\tau=1/t$. It can be
easily seen that, in going from the open (closed)
to the closed (open) string channel, we have the following correspondence
between the various non vanishing spin structures~\cite{DL99121}:
\begin{eqnarray}
&& {\rm NS} \leftrightarrow  {\rm NS-NS}~, ~~~~~~~~{\rm NS (-1)^F}
\leftrightarrow {\rm R-R}~~\nonumber\\
&& R \leftrightarrow  {\rm NS-NS (-1)^F}~,~~{\rm R
(-1)^F} \leftrightarrow {\rm R-R (-1)^F} .
\label{spinstruco}
\end{eqnarray}
It is also easy to see that the distance $y$ between the dressed
D3 brane and the stack of the $N$ D3 branes  makes the integral in
Eq.~(\ref{closed}) convergent for small values of $t$, while in
the limit $t \rightarrow \infty$, the integral is logarithmically
divergent. This divergence is  due to a twisted tadpole corresponding
to the exchange of massless closed string states between the two
boundary states in Eq. (\ref{zg32}). We would like to stress that
the presence of the gauge field
makes the divergence to appear already {\em at the string level}, before
any field theory limit ($\alpha ' \rightarrow 0$) is performed.
When $F$ vanishes, the divergence is eliminated by the integrand
being identically zero as a consequence of the fact that
fractional branes are BPS states.

As observed in Refs.~[10 $\div$ 13, 19] tadpole
divergences  correspond in general to the presence of gauge anomalies,
which make the gauge theory inconsistent and must be eliminated by
drastically modifying the theory or by fixing particular values of
parameters. For instance, in type I superstring they are eliminated by
fixing the gauge group to be $SO(32)$~\cite{PC}. Instead, as stressed in
Refs.~[10 $\div$ 13], logarithmic tadpole divergences do
not correspond to gauge anomalies. In the bosonic string
they have been cured in a variety of
ways~[14 $\div$ 20]. It will turn out in the next Section
that, in our case, the logarithmically
divergent tadpoles correspond to the fact that the gauge theory living
on the brane is not conformally invariant. In fact, they
provide the correct one-loop running coupling constant.

In this paper, following the suggestion of Refs.~\cite{LR,KAKURO}, we
cure these divergences just by introducing
in Eq. (\ref{closed}) an infrared cutoff that regularizes the
contribution of the massless closed string states. Since, in the open/closed
string duality, an infrared divergence in the closed string channel
corresponds to an ultraviolet divergence in the open string
channel, it is easy to see that the expression in
Eq. (\ref{zetatet}) is divergent for small values of $\tau$ and needs
an ultraviolet cutoff. It will turn out that this divergence is exactly
the one-loop divergence that one gets in ${\cal{N}}=2$ super
Yang-Mills, which is the
gauge theory living in the world-volume of the fractional D3 brane.

Our results are consistent with those of Ref.~\cite{BIANCHI}
where, in the context of unoriented theories, it is shown that
string amplitudes are in general affected by both ultraviolet and
infrared divergences. In our case, the finite parameter $y$
provides  a natural IR (UV) cut-off in the open (closed) channel,
leaving the amplitude to be divergent only in the UV (IR) corner.
Following their procedure, we could as well introduce two stringy
$\beta$-functions, one for each channel, and show that they
coincide. However we are interested in establishing a deeper
connection with the gauge/gravity correspondence and therefore in
this paper we focus on the behavior of massless and massive states
separately with respect to the open/closed duality.

To this end, starting from
Eqs.~(\ref{zetatet}) and ~(\ref{closed}), containing arbitrary powers of the
gauge field $F$, we extract  the quadratic term in the gauge
field $F$ from the previous general expressions.
In the open string channel it is given by:
\begin{eqnarray}
Z_h^o(F)\!\!&=&\!\!\left[- \frac{1}{4} \int d^4 x
F_{\alpha \beta}^{a} F^{a\,\alpha \beta}
\right] \left\{
 - \frac{N}{8 \pi^2} \int_{0}^{\infty}
\frac{d \tau}{\tau}
{e}^{-\frac{y^{2} \tau}{2 \pi \alpha' } }
\left[\frac{f_3(k)f_4(k)}{f_1(k)f_2(k)}\right]^4
2k\frac{d}{dk} \log\left[\frac{f_3(k)}{f_4(k)}\right]
\right\}\nonumber\\
&& + iN \left[\frac{1}{32\pi^2} \int d^4x
F^a_{\alpha\beta}\tilde F^{a\,\alpha\beta} \right] \int_{0}^{\infty}
\frac{d\tau}{\tau}e^{-\frac{y^{2} \tau}{2\pi\alpha'}}
~,~~~~~ k = {e}^{- \pi \tau}~~,
\label{gau56}
\end{eqnarray}
while in the closed string channel it reads:
\begin{eqnarray}
Z_h^c(F)\!\!&=&\!\!\left[- \frac{1}{4} \int d^4 x F_{\alpha \beta}^{a}
F^{a\,\alpha \beta }
\right] \left\{
   \frac{N}{8 \pi^2}
\int_{0}^{\infty} \frac{dt}{t}  {e}^{- \frac{y^{2}}{2 \pi\alpha' t} }
\left[\frac{f_3(q)f_2(q)}{f_1(q)f_4(q)}\right]^4
2q\frac{d}{dq} \log\left[\frac{f_3(q)}{f_2(q)}\right] \right\}
\nonumber\\
&& + iN \left[ \frac{1}{32\pi^2} \int d^4x
F^a_{\alpha\beta}\tilde F^{a\,\alpha\beta} \right]
\int_{0}^{\infty} \frac{dt}{t}
e^{-\frac{y^{2} }{2\pi\alpha't}}
~,~~~~~ q = {e}^{-\pi t}~~.
\label{F289}
\end{eqnarray}
The two previous equations are derived in
detail in Appendices \ref{app2} and \ref{app4} respectively, and are
equal to
each other as one can see by performing  the modular transformation
$\tau = 1/t$. It can also be seen that the untwisted sector
does not produce any quadratic term in the gauge field,
in analogy with what happens in flat space because of its
non-renormalization in ${\cal{N}}=4$ super Yang-Mills~\footnote{For
  details see for instance Ref.~\cite{BF} where also the orbifold
$\mathbb{C}^2 /\mathbb{Z}_2$ is considered in the compactification of
type I superstring.}.

As we have already discussed, it turns out that the divergence that we
have at the string level is exactly the one-loop divergence present in the
gauge field theory living in the world-volume of the brane and can be
equivalently seen as due, in the open string channel, to the massless
open string states circulating in the loop and, in the closed string
channel, to the massless closed string states exchanged between two
boundary states. This can be seen by
isolating the contribution of the massless open and closed string
states, respectively, in Eqs.~(\ref{gau56}) and ~(\ref{F289}), which
are the only ones giving a divergence. In this way in the open
string channel we get:
\begin{eqnarray}
Z_h^o(F)\!\!&\rightarrow&\!\!\left[- \frac{1}{4} \int d^4 x F_{\alpha
\beta}^{a} F^{a\,\alpha \beta } \right] \nonumber \\
& \times &\!\!
\left\{
\frac{1}{g_{YM}^{2} (\Lambda )}
 - \frac{N}{8 \pi^2}
\int_{ {\small \frac{1}{\alpha' \Lambda^2}} }^{\infty} \frac{d \tau}{\tau}
{e}^{-\frac{y^{2} \tau}{2 \pi \alpha' } } + \frac{N}{8 \pi^2}
\int_{0}^{\infty} \frac{d \tau}{\tau}
{e}^{-\frac{y^{2} \tau}{2 \pi \alpha' } } G (k) \right\} \nonumber\\
&+&  iN \left[\frac{1}{32\pi^2} \int d^4x F^a_{\alpha\beta}\tilde
F^{a\,\alpha\beta} \right] \int_{  {\small \frac{1}{\alpha'
    \Lambda^2}} }^{\infty}
\frac{d\tau}{\tau}e^{-\frac{y^{2} \tau}{2\pi\alpha'}}
\label{gau57}
\end{eqnarray}
where
\begin{eqnarray}
G (k) = -\left[\frac{f_3(k)f_4(k)}{f_1(k)f_2(k)}\right]^4
2k\frac{d}{dk} \log\left[\frac{f_3(k)}{f_4(k)}\right] +1,
\label{Gk89}
\end{eqnarray}
while in the closed string channel we get:
\begin{eqnarray}
Z_h^c(F)\!\!&\rightarrow&\!\!\left[- \frac{1}{4} \int d^4 x
F_{\alpha \beta}^{a}
F^{a\,\alpha \beta }\right] \nonumber \\
& \times & \! \!
 \left\{
\frac{1}{g_{YM}^{2}(\Lambda ) }
- \frac{N}{8 \pi^2}
\int_{0}^{\alpha' \Lambda^2}
\frac{dt}{t}  {e}^{- \frac{y^{2}}{2 \pi\alpha' t} }
+\frac{N}{8 \pi^2}
\int_{0}^{\infty} \frac{dt}{t}  {e}^{- \frac{y^{2}}{2 \pi\alpha' t} }
F(q) \right\}  \nonumber\\
&+&  iN \left[ \frac{1}{32\pi^2} \int d^4x
F^a_{\alpha\beta}\tilde F^{a\,\alpha\beta} \right]
\int_{0}^{ {\alpha' \Lambda^2 }} \frac{dt}{t}
e^{-\frac{y^{2} }{2\pi\alpha't}}
\label{F285}
\end{eqnarray}
where
\begin{eqnarray}
F(q)= \left[\frac{f_3(q)f_2(q)}{f_1(q)f_4(q)}\right]^4
2q\frac{d}{dq} \log\left[\frac{f_3(q)}{f_2(q)}\right]  + 1~~.
\label{Fq89}
\end{eqnarray}
Notice that in the two previous equations we have also added the
contribution coming from the tree diagrams that contain the bare
coupling constant. In an ultraviolet finite theory such as string theory we
should not deal with a bare and a renormalized coupling. On the other
hand, we have already discussed the fact that the introduction of a
gauge field produces a string amplitude that is divergent already at the
string level and that therefore must be regularized with the
introduction of a cutoff.

We have already mentioned that Eqs. (\ref{gau57}) and (\ref{F285}) are
equal to each other as one can see by performing the modular
transformation $\tau = 1/t$. Actually, by a closer look one can
see that the contribution of the massless states and the one of the
massive states transform respectively into each other without any mixing
between massless and massive states. Indeed the contribution of the massless 
closed
string states can be easily obtained
by performing the modular transformation on that of the massless open
string states and viceversa. Furthermore
the threshold corrections, corresponding to the
contribution of the massive states in the two channels, are
exactly  equal as a consequence of the following equation:
\begin{eqnarray}
F (q) = G (k)
\label{FqGk}
\end{eqnarray}
that can be easily proven using the modular transformations of the
functions $f_i$ given in Appendix~\ref{app0} and Eq. (\ref{kqtau}).

This means that the open/closed
string duality exactly maps the ultraviolet divergent contribution
coming from the
massless open string states circulating in the loop - and that reproduces
the divergences of ${\cal{N}}=2$ super Yang-Mills living in the
world-volume of the fractional D3 branes - into the infrared divergent
contribution due to the massless closed string states
propagating between the two boundary states.
This leads to the first evidence why the one-loop running
coupling constant
can be consistently derived from a supergravity calculation as
originally shown in Refs.~[21 $\div$ 23] and reviewed in
Ref.~\cite{REV}. This will be shown in a more direct and quantitative
way in the next Section.

\section
{Field Theory Limit in the Two Channels}
\label{sez2}

In this Section
 we perform the field theory limit of
the amplitudes given by Eqs.~(\ref{gau57}) and (\ref{F285})  in
the open and closed string channel, respectively.
In both channels the
field theory
limit is obtained by performing the zero slope limit
($\alpha'\rightarrow 0$) {\em  together} with the limit in which
the modular variables $t$ and $\tau$ go to infinity, keeping fixed
the dimensional Schwinger proper times  $\sigma
 = \alpha' \tau$
and  $s = \alpha' t$  of the open and closed string, respectively. Indeed,
in these two limits the only surviving contributions in
Eqs.~(\ref{gau57}) and~(\ref{F285})  are those due to the massless
states.
Notice that the two regions $t\rightarrow \infty$ and
$\tau\rightarrow\infty$ are not connected to each other through a modular
transformation.

Let us start performing the field theory limit in the open string
channel as explained above, namely by taking $\tau \rightarrow
\infty$, $\alpha ' \rightarrow 0$ with $\alpha' \tau \equiv
\sigma$ fixed. In so doing we see that in Eq.~(\ref{gau57}) the
contribution of the massive open string states vanishes and we get:
\begin{eqnarray}
Z_h^o(F)\!\!\!&& \rightarrow
\left[ - \int d^4x\frac{1}{4} F_{\alpha \beta}^a
F^{a\, \alpha \beta} \right]
\left[ \frac{1}{g_{YM}^{2} (\Lambda) } - \frac{N}{8\pi^2}
\int_{1/\Lambda^2}^{\infty} \frac{d\sigma}{\sigma}
e^{-\frac{y^{2} \sigma}{2\pi (\alpha')^2}} \right]
\nonumber\\
\!\!\!&& + iN \left[\frac{1}{32\pi^2}\int d^4x
F_{\alpha\beta}^a \tilde F^{a\,\alpha\beta}\right]
\int_{1/\Lambda^2}^{\infty} \frac{d\sigma}{\sigma}
e^{-\frac{y^{2} \sigma}{2\pi (\alpha')^2}} \, .
\label{opftlim}
\end{eqnarray}
The previous integrals are naturally
regularized in the infrared regime ($\sigma \rightarrow \infty$) by the
fact that the two stacks of branes are at a finite distance $y$.
Notice that, in order to get a finite expression in the field
theory limit, we also need to take the limit $y \rightarrow 0$ while
keeping fixed the quantity $\frac{y}{\alpha'}$. This finite quantity
is directly related to the complex scalar field $\Psi$ of the
${\cal{N}}=2$ gauge supermultiplet by the gauge/gravity relation~\cite{CJ},
\begin{equation}
\Psi= \frac{y}{2\pi\alpha'}e^{i\theta} \qquad {\rm with} \qquad
x^4  + i x^5 \equiv y e^{i\theta} \, ,
\label{ysc}
\end{equation}
and the fact that $\Psi$ has a nonzero vacuum expectation value does
not enlarge the gauge group $SU(N)$.

If we perform the field theory limit in the closed string
channel by taking $t \rightarrow \infty$, $\alpha' \rightarrow 0$
with $ \alpha ' t \equiv s$ fixed, we get from Eq. (\ref{F285}):

\begin{eqnarray}
Z_h^c(F)\!\!\!&& \rightarrow \left[ - \int d^4x\frac{1}{4} F_{\alpha \beta}^a
F^{a\, \alpha \beta} \right]
\left[ \frac{1}{g_{YM}^{2} (\Lambda)} - \frac{N}{8\pi^2}
\int_{0}^{(\alpha'\Lambda)^2} \frac{d s}{s}
e^{-\frac{y^{2} }{2\pi s}}\right]
\nonumber\\
\!\!\!&& + iN \left[\frac{1}{32\pi^2}\int d^4x
F_{\alpha\beta}^a \tilde F^{a\,\alpha\beta}\right]
\int_{0}^{(\alpha'\Lambda)^2}\frac{d s}{s}
e^{-\frac{y^{2} }{2\pi s}}  \, .
\label{clftlim}
\end{eqnarray}

In the closed string case the distance $y$ between the branes  makes the
integral convergent in the ultraviolet regime ($s \rightarrow 0$), but
instead  an infrared cutoff $\Lambda$ is needed.
If we identify the two $\Lambda$'s appearing in
the ultraviolet cutoff in the open string channel and in the
infrared cutoff in the closed string one, we see that the
expressions  in the two field theory limits are actually
equal. This observation clarifies now why the supergravity
solution gives the correct answer for the perturbative behavior
of the non-conformal world-volume theory as found in
Refs.~[21 $\div$ 25].
In fact we can extract the coefficient of the term $F^2$ from either
of the two Eqs.~(\ref{opftlim}) and ~(\ref{clftlim}) obtaining the
following expression:

\begin{eqnarray}
\frac{1}{g_{YM}^{2} (\epsilon) } + \frac{N}{8\pi^2} \log
\frac{y^{2}}{\epsilon^2}  \equiv \frac{1}{g_{YM}^{2} (y ) }~~,~~
\epsilon^2 \equiv 2 \pi (\alpha' \Lambda )^2
\label{run2}
\end{eqnarray}
where the integral appearing in Eq.~(\ref{opftlim}) has been explicitly
computed:

\begin{eqnarray}
I (\Lambda,y) \equiv \int_{1/\Lambda^2}^{\infty}
\frac{d\sigma}{\sigma} e^{-\frac{y^2 \sigma}{2\pi (\alpha')^2}}
\simeq \log \frac{2 \pi
  (\alpha' \Lambda)^2}{y^2}
\label{comple67}
\end{eqnarray}
Eq. (\ref{run2}) exactly reproduces Eq. (143) of Ref.~\cite{REV}, with
$N_2 =0, N_1 =N$. In that paper
the renormalization group parameter $\mu$ that describes the running is
related to the distance $y$ between the two stacks of branes
precisely by the relation $\mu = \frac{y}{2 \pi \alpha'}$, while
$\epsilon$ is an ultraviolet cutoff. The previous derivation makes it
clear why the running coupling constant of ${\cal{N}}=2$ super
Yang-Mills can be obtained from the supergravity solution
corresponding to a fractional D3 brane of the orbifold
$\mathbb{C}^2 /\mathbb{Z}_2$.

Eq.~(\ref{run2}) gives the one-loop correction to the bare gauge
coupling constant $g_{YM} (\Lambda )$ in the gauge
theory regularized with cutoff $\Lambda$. The renormalization
procedure can then be performed by introducing the renormalized coupling
constant $g_{YM} (\mu)$ related to the bare one by:
\begin{eqnarray}
 \frac{1}{g_{YM}^{2} (\Lambda)} =
\frac{1}{g_{YM}^{2} (\mu) } + \frac{N}{8\pi^2} \log
\frac{\Lambda^{2}}{\mu^2}
\label{run23}
\end{eqnarray}
being $\mu$ the renormalization scale.
Using Eq. (\ref{run23}) in either Eq. (\ref{opftlim}) or Eq. (\ref{clftlim})
one gets the following expression for the coefficient of the $F^2$
term
\begin{eqnarray}
 \frac{1}{g_{YM}^{2} (\mu)}  - \frac{N}{8\pi^2} \log
\frac{\mu^{2}}{m^2} = \frac{1}{g_{YM}^{2} (m)}~~~;~~~
m^2 \equiv \frac{y^2}{2 \pi \alpha^{'2}} .
\label{run24}
\end{eqnarray}
From it, or equivalently from Eq. (\ref{run23}), we can now determine the
one-loop $\beta$-function:
\begin{eqnarray}
\beta (g_{YM}) \equiv \mu \frac{\partial}{\partial \mu} g_{YM}(\mu) =
-\frac{g^{3}_{YM} N}{8 \pi^2}   \label{betafun}
\end{eqnarray}
which is the correct one for ${\cal{N}}=2$ super Yang-Mills.

Let us turn now to the vacuum angle $\theta_{YM}$, provided
by the terms in Eqs.~(\ref{opftlim}) and ~(\ref{clftlim}) with the
topological charge. If we extract it from either of these two equations
we find that it is imaginary and,
moreover, must be renormalized like the coupling constant.
A way of eliminating these problems is
to introduce a complex cutoff and to allow
the gauge field to be in either
one of the two sets of branes by taking the symmetric
combination~\footnote{We thank M. Bill\'o, F. Lonegro and I. Pesando for
useful discussions on this point.}:
\begin{eqnarray}
\frac{1}{2} \left[\langle D3; F | D  |D3 \rangle +
\langle D3 | D  | D3; F \rangle \right] =
\frac{1}{2} \left[\langle D3; F | D  |D3 \rangle +
\overline{\langle D3; F | D  | D3 \rangle} \right].
\label{sym78}
\end{eqnarray}
If we introduce a complex cutoff $ \Lambda \rightarrow \Lambda {e}^{-i
  \theta} $ the divergent
integral in Eq.~(\ref{comple67}) becomes:
\begin{eqnarray}
I (z) \equiv \int_{1/\Lambda^2}^{\infty}  \frac{d\sigma}{\sigma}
e^{-\frac{y^2 \sigma}{2\pi (\alpha')^2}} \simeq \log \frac{2 \pi
  (\alpha')^2 \Lambda^2}{y^2 {e}^{ 2i \theta}}~.
\label{comple68}
\end{eqnarray}
This procedure leaves unchanged all the previous considerations concerning
the gauge coupling constant, because in this case the coefficient of
the $F^2$ term results to be proportional to the
following combination:
\begin{eqnarray}
\frac{1}{2} \left[ I (z) + I ({\bar{z}}) \right] \simeq \log \frac{2 \pi
  (\alpha')^2 \Lambda^2}{y^{2}}~~.
\label{gau73}
\end{eqnarray}
In the case of the $\theta_{YM}$ angle one gets instead:
\begin{eqnarray}
\theta_{YM} = i\frac{N}{2} \left [ I (z) - I ({\bar{z}}) \right] =
2 N\theta~~~,
\label{theta45}
\end{eqnarray}
exactly reproducing the result given in Eq. (144) of
Ref.~\cite{REV} if we take again $N_2 = 0$ and $ N_1 =N$.
Remember, however, that in Ref.~\cite{REV} $\theta$ is the phase
of the complex quantity $z = y {e}^{i \theta}$. But, as it can be
seen in Eq. (\ref{comple68}), giving a phase to the cutoff
corresponds to giving the opposite phase to the distance between
the branes $y$. We prefer to make the cutoff complex rather than
$y$ in order to keep the open string Virasoro generator $L_0$
real.

\section
{{\bf{Branes in External Fields: ${\cal N} =1$ orbifold}}}
\label{sez4}

In the following we extend the analysis performed in the previous
Section
to the case of the orbifold $\mathbb{ C}^3/(\mathbb{ Z}_2\times
\mathbb{Z}_2)$ preserving four supersymmetry charges. The orbifold group
$\mathbb{ Z}_2\times\mathbb{Z}_2$ contains four elements whose action on
the three complex coordinates
\begin{eqnarray}
&&z_1=x_4+ix_5 \qquad z_2=x_6+ix_7 \qquad z_3=x_8+ i x_9
\label{z123}
\end{eqnarray}
is chosen to be
\begin{eqnarray}
R_e= \left[
\begin{array}{ccc}
~~1 & ~0& ~~0
\\ ~~0& ~1 & ~~0 \\ ~~0 & ~0  & ~~1
\end{array}
\right],  && \quad  R_{h_1}=\left[
\begin{array}{ccc}
1 & 0& 0
\\ 0& -1 & 0 \\ 0 & 0  & -1
\end{array}
\right], \nonumber\\
 \quad R_{h_2}= \left[
\begin{array}{ccc}
-1 & 0& 0
\\ 0& 1 & 0 \\ 0 & 0  & -1
\end{array}
\right], &&\quad   R_{h_3}= \left[
\begin{array}{ccc}
-1 & 0& 0
\\ 0& -1 & 0 \\ 0 & 0  & 1
\end{array}
\right] \ .
\end{eqnarray}

As previously stated, fractional
branes have Chan-Paton factors transforming according to irreducible
representations of the orbifold group. The group
$\mathbb{ Z}_2\times \mathbb{ Z}_2$ has four
irreducible
one-dimensional representations corresponding to four different
kinds of  fractional branes. The orbifold $\mathbb{
C}^3/(\mathbb{ Z}_2\times \mathbb{ Z}_2)$, as already explained
Refs.~\cite{ NAPOLI,FERRO}, can be seen as obtained by three copies of
the orbifold
$\mathbb{ C}^2/\mathbb{ Z}_2$ where the i-th $\mathbb{Z}_2$ contains
the elements $(e, h_i)$
$(i=1, \dots 3)$. This means that the boundary states associated to each
fractional brane are:
\begin{eqnarray}
&&| Dp>_1 =| Dp>_u+ | Dp>_{t_1}+ | Dp>_{t_2} +| Dp>_{t_3} \ , \nonumber\\
&&| Dp>_2 =| Dp>_u+ | Dp>_{t_1}- | Dp>_{t_2} -| Dp>_{t_3} \ ,\nonumber \\
&&| Dp>_3 =| Dp>_u- | Dp>_{t_1}+ | Dp>_{t_2} -| Dp>_{t_3} \ ,\nonumber \\
&&| Dp>_4 =| Dp>_u- | Dp>_{t_1}- | Dp>_{t_2} +| Dp>_{t_3} \ ,\label{bs}\\
\nonumber
\end{eqnarray}
where $| Dp>_u$ is the untwisted boundary state that, apart an overall factor
$\frac{1}{2}$ due to the orbifold projection, is the same as the one
in flat space, and
$ | Dp>_{t_i}$ $(i=1, \dots, 3)$ are exactly the same as the twisted
boundary states on the orbifold $\mathbb{ C}^2/\mathbb{Z}_2$, apart
from a factor
$1/\sqrt 2$. The signs in front of each twisted term in Eq. (\ref{bs})
depend on the irreducible representation chosen for the orbifold group action
on the Chan-Paton factors.

In order to keep the forthcoming discussion as general as
possible, we study the interaction between a stack of $N_I$ ($I=1,
\dots, 4$) branes of type $I$ and a D3-fractional brane, for
example of type $I=1$, with an $SU(N)$ gauge field turned on its
world-volume. Due to the structure of the orbifold  $\mathbb{
C}^3/(\mathbb{ Z}_2\times \mathbb{ Z}_2)$, this interaction is the
sum of four terms:
\begin{eqnarray}
Z= Z_{e} + \sum_{i=1}^{3} Z_{h_i}~~,    \label{ZN=1}
\end{eqnarray}
where $Z_e$ and $Z_{h_i}$ are obtained in the open [closed]  channel
by multiplying the Eqs.~(\ref{zetae}) and (\ref{zetatet}) [Eqs.~(\ref{op2})
and (\ref{closed})] by an extra 1/2 factor due to the orbifold projection.
Also in this case we limit our considerations to the twisted sectors,
which are the only ones that provide a non-zero contribution to the
quadratic terms in the gauge field. Their contribution, in the open
string channel, is given by:

\begin{eqnarray}
Z_{h_i}^o & = & \frac{f_i(N)}{2\,(8\pi^2\alpha')^2} \int d^4x
\sqrt{ -\mbox{det}(\eta+\hat{F})} \int_{0}^\infty
\frac{d \tau}{\tau}e^{-\frac{y^{2}_i \tau }{2\pi\alpha'}}
\frac{4\,\sin\pi\nu_f \sin\pi\nu_g}{ \Theta_{2}^2(0|i \tau)
\Theta_{1}(i \nu_f \tau| i \tau) \Theta_{1}(i \nu_g \tau|i \tau)}
\nonumber\\
&&\times \left\{\Theta_{3}^2(0|i \tau)
\Theta_{4}(i \nu_f \tau|i \tau)\Theta_{4}(i \nu_g \tau |i \tau)
-\Theta_{4}^2(0|i \tau) \Theta_{3}(i \nu_f \tau|i \tau)
\Theta_{3}(i \nu_g \tau |i \tau)\right\}\nonumber \\
&&+\frac{i\, f_i(N)}{64\pi^2} \int d^4x F^a_{\alpha\beta}\tilde
F^{a \, \alpha\beta}\int \frac{d\tau}{\tau} e^{-\frac{y_i^{2} \tau }{2\pi\alpha'}}~~.
\label{ztet1}
\end{eqnarray}
In the previous expression we should put to zero the distance
$y_i$ between the stack of the $N_I$'s branes and the dressed one, since
the fractional branes are constrained to live at the
orbifold fixed point $z_1=z_2=z_3=0$. However, $y_i$
provides a natural infrared cutoff in Eq.~(\ref{ztet1}).
Therefore we keep this quantity small but finite, and we will put
it to zero just at the end of the calculation.
The functions $f_i(N)$ introduced in Eq.~(\ref{ztet1}) depend on
the number of the different kinds of fractional branes $N_I$, and their
explicit expression is given by~\cite{NAPOLI,FERRO}:
\begin{eqnarray}
&& f_1(N_I)= N_1 + N_2 - N_3 - N_4 \, , \nonumber\\  && f_2(N_I)= N_1 -
N_2 + N_3 - N_4 \, , \nonumber\\  && f_3(N_I)= N_1 - N_2 - N_3 + N_4 \, .
\label{coupling}
\end{eqnarray}
One can follow the same steps in the closed string channel obtaining
the formulas corresponding to Eq. (\ref{closed}).

Let us now extract in both channels the quadratic terms in the gauge
field $F$. In the open sector, we get:
\begin{eqnarray}
Z_h^o(F)\!\!&\rightarrow&\!\!\left[- \frac{1}{4} \int d^4 x
F_{\alpha \beta}^{a} F^{a\,\alpha \beta } \right] \nonumber \\
& \times & \!  \! \left\{
\frac{1}{g_{YM}^{2} (\Lambda )}
 - \sum_{i=1}^{3} \frac{f_i(N) }{16 \pi^2} \left[
\int_{1/(\alpha' \Lambda^2)}^{\infty} \frac{d \tau}{\tau}
{e}^{-\frac{y^{2}_i \tau}{2 \pi \alpha' } } +
\int_{0}^{\infty} \frac{d \tau}{\tau}
{e}^{-\frac{y^{2}_i \tau}{2 \pi \alpha' } } G (k) \right] \right\} \nonumber\\
& + & i \left[\frac{1}{32\pi^2} \int d^4x F^a_{\alpha\beta}\tilde
F^{a\,\alpha\beta} \right] \sum_{i=1}^3 \frac{f_i(N)}{2}
\int_{\frac{1}{\alpha'\Lambda^2}}^{\infty}
\frac{d\tau}{\tau}e^{-\frac{y^{2}_i \tau}{2\pi\alpha'}}~~,
\label{gau571}
\end{eqnarray}
while in the closed string channel we obtain:
\begin{eqnarray}
Z_h^c(F)\!\!&\rightarrow&\!\!\left[- \frac{1}{4} \int d^4 x
F_{\alpha \beta}^{a} F^{a\,\alpha \beta }\right] \nonumber \\
& \times & \! \! \left\{
\frac{1}{g_{YM}^{2}(\Lambda ) } - \sum_{i=1}^{3} \frac{f_i(N)}{16 \pi^2}
\left[
\int_{0}^{\alpha' \Lambda^2} \frac{dt}{t}  {e}^{- \frac{y^{2}_i}{2
\pi\alpha' t} } + \int_{0}^{\infty}
\frac{dt}{t} {e}^{- \frac{y^{2}_i}{2 \pi\alpha' t} }
F(q) \right] \right\}  \nonumber\\
&+&  i \left[ \frac{1}{32\pi^2} \int d^4x F^a_{\alpha\beta}\tilde
F^{a\,\alpha\beta} \right] \sum_{i=1}^{3} \frac{f_i(N)}{2}
\int_{0}^{\alpha' \Lambda^2} \frac{dt}{t}
e^{-\frac{y_i^{2} }{2\pi\alpha't}} ~~. \label{F2851}
\end{eqnarray}
Analogously to the case of the ${\cal N}=2$ orbifold, we have
isolated, in both channels, the divergent contribution due to the
massless states, and we have also added the one
coming from the tree diagrams. The main properties exhibited by the
interactions in the orbifold $\mathbb{C}^2 /\mathbb{Z}_2$ in
Eqs.~ (\ref{gau57}) and (\ref{F285})
are also shared by the interactions in the orbifold
$\mathbb{C}^3/(\mathbb{Z}_2 \times \mathbb{Z}_2)$ in
Eqs.~(\ref{gau571}) and
(\ref{F2851}). In particular, also in this case,
one can see that the contribution of the massless states and that  of the
massive states transform respectively into each other without any mixing
between them. This confirms the main result obtained in Sect. 2,
i.e. the open/closed string duality
exactly maps the ultraviolet divergent contribution coming from the massless
open string states - which in this case reproduces the divergences of
${\cal N}=1$ super Yang-Mills - into the infrared divergent contribution
due to the massless closed string states.

Extracting the coefficient of the term $F^2$ from Eq. (\ref{gau571}) or
(\ref{F2851}), and performing on it the field theory limit as
explained in the previous Section
, we get:
\begin{eqnarray}
\frac{1}{g^2_{\rm YM} (\epsilon)}+ \frac{1}{8\pi^2} \sum_{i=1}^3
f_i(N_I) \log\frac{y_i}{\epsilon}\equiv
\frac{1}{g^2_{\rm YM}}(y)\qquad
\epsilon^2= 2\pi (\alpha' \Lambda)^2~~.
\label{run21}
\end{eqnarray}
Eq. (\ref{run21}) reproduces Eq. (3.14) of  Ref.~\cite{ANOMA}, and
this explains again why the supergravity solutions, dual to ${\cal
N}=1$ super Yang-Mills theory, give the correct answer for the
perturbative behavior of the non conformal world-volume theory, as
found in Refs.~\cite{ANOMA, NAPOLI, FERRO}.

Performing the renormalization procedure, we introduce the
renormalized
coupling constant $g_{YM}(\mu)$
given in terms of the bare one by the relation:
\begin{eqnarray}
 \frac{1}{g_{YM}^{2} (\Lambda)} =
\frac{1}{g_{YM}^{2} (\mu) } + \sum_{i=1}^{3}\frac{f_i(N)}{16\pi^2} \log
\frac{\Lambda^{2}}{\mu^2}=
\frac{1}{g_{YM}^{2} (\mu) } + \frac{3N_1 - N_2 -N_3 -N_4}{16\pi^2} \log
\frac{\Lambda^{2}}{\mu^2}~~.
\label{run231}
\end{eqnarray}
From this equation we can determine the $\beta$-function obtaining:
\begin{eqnarray}
\beta (g_{YM}) \equiv \mu \frac{\partial}{\partial \mu} g_{YM}(\mu) =
-\frac{g^{3}_{YM}}{16 \pi^2} \, \, (3N_1 - N_2 - N_3 -N_4) \, ,
\label{betafun1}
\end{eqnarray}
that is  the correct one for the wold-volume theory living on the
dressed brane.

Finally, in the same spirit as in Sect. 3, we  consider the symmetric
combination given in  Eq.~(\ref{sym78})
and, by introducing a complex cut-off $\Lambda e^{-i\theta}$, or
equivalently, considering
complex coordinates $z_i= y_i e^{i\theta}$, we arrive to the
following expression for the $\theta_{YM}$:
\begin{eqnarray}
\theta_{YM}= \sum_{i=1}^3 f_i(N_I)\theta \,
\label{theta451}
\end{eqnarray}
again in agreement with the result given in Eq.~(3.14) of the
Ref.~\cite{ANOMA}.

\section
{Conclusions}

In this paper we have used open/closed string duality of the annulus
diagram for explaining
why the perturbative properties of ${\cal N}=1,2$ Super Yang-Mills
theories, living in the world-volume of fractional D3 branes, follow
directly from their classical supergravity
solutions~[21 $\div$ 25]. This is a
consequence of the fact that the coefficient of the gauge kinetic term
in the annulus diagram is expressed by an integral that is divergent
already at the string
level and, therefore, must be regularized. It turns out that this
divergent contribution can be seen to be equivalently due either to the
exchange of massless closed strings between two boundary states in the
closed string channel or to the massless open string states
circulating in the loop in the open string channel, and that these two
contributions are mapped into each other by the modular transformation that
relates the open and closed string channels. This means that the
divergence present at the string level is precisely the one that
one gets in the gauge field theory living on the brane in the field
theory limit ($\alpha' \rightarrow 0$). It remains unclear why these
two kinds of divergences must be related. This is in fact not the case,
for instance, of the tadpoles present in the bosonic
string. Here in fact we have a dilaton tadpole that is  not
related to the divergences due to the massless gauge fields
circulating in the loop in pure Yang-Mills theory for $d=26$, which
are  proportional to $d-26$~\cite{FILIM,MT} and therefore vanish
for $d=26$. Supersymmetry may play a role in the identification of the
two divergences~\cite{BIANCHI}. It must also be noticed that the situation 
here is
different from what was found for the gauge kinetic term in the case of
the heterotic string~\cite{K}, where no ultraviolet, but only
infrared divergences are found at the string level and these
divergences are due to the contribution of the massless string states
that are also present in the limiting field theory.

{\bf{Acknowledgments}}
We deeply thank M. Bill\'{o}, F. Lonegro and I. Pesando for
useful discussions. We also thank W. Mueck, R. Musto, F. Nicodemi, R. Pettorino
for reading the manuscript. F. Pezzella thanks Nordita for their kind
hospitality.

\appendix

\section
{$\Theta$ functions}
\label{app0}

The $\Theta$-functions which are the solutions of  the heat
equation
\begin{eqnarray}
\frac{\partial}{\partial t}
\Theta\left(\nu|i t\right)=\frac{1}{4\pi} \partial_{\nu}^2
\Theta(\nu|i t) \label{iden}
\end{eqnarray}
are given by:
\begin{eqnarray} &&\Theta_1(\nu
|it)\equiv\Theta_{11}(\nu,|it) =-2 q^{\frac{1}{4}}\sin\pi\nu
\prod_{n=1}^\infty \left[(1-q^{2n}) (1-e^{2i\pi\nu}q^{2n})
(1-e^{-2i\pi\nu}q^{2n}) \right]~~,
\nonumber\\
&&\Theta_2(\nu |it)\equiv\Theta_{10}(\nu,|it)
=2 q^{\frac{1}{4}}\cos\pi\nu \prod_{n=1}^\infty \left[(1-q^{2n})
(1+e^{2i\pi\nu}q^{2n})(1+e^{-2i\pi\nu}q^{2n}) \right]~~,
\nonumber\\
&&\Theta_3(\nu,|it)\equiv\Theta_{00}(\nu,|it)
=\prod_{n=1}^\infty \left[(1-q^{2n})(1+e^{2i\pi\nu}q^{2n-1})
(1+e^{-2i\pi\nu}q^{2n-1}) \right]~~,
\nonumber\\
&&\Theta_4(\nu,|it)\equiv\Theta_{01}(\nu,|it)
=\prod_{n=1}^\infty \left[(1-q^{2n})(1-e^{2i\pi\nu}q^{2n-1})
(1-e^{-2i\pi\nu}q^{2n-1}) \right]~~,
\end{eqnarray}
with $q=e^{-\pi t}$.
The modular transformation properties of the $\Theta$ functions are
\begin{eqnarray}
&&\Theta_1(\nu |it)
=i\,\Theta_{1}(-i\frac{\nu}{t} |\frac{i}{t})e^{-\pi\nu^2 /t} t^{-\frac{1}{2}}~~,
\nonumber\\
&&\Theta_{2,\,3,\,4}(\nu|it)
=\Theta_{4,\,3,\,2}(-i\frac{\nu}{t}|\frac{i}{t})e^{-\pi\nu^2/t}
t^{-\frac{1}{2}}~~.
\label{modtras}
\end{eqnarray}

It is also useful to define the $f$-functions and their transformation
properties. We define
\begin{eqnarray}
  \label{f1}
  f_1\equiv q^{{1\over 12}} \prod_{n=1}^\infty (1 - q^{2n})~~,
\end{eqnarray}
\begin{eqnarray}
  \label{f2}
  f_2\equiv \sqrt{2}q^{{1\over 12}} \prod_{n=1}^\infty (1 + q^{2n})~~,
\end{eqnarray}
\begin{eqnarray}
  \label{f3}
  f_3\equiv q^{-{1\over 24}} \prod_{n=1}^\infty (1 + q^{2n -1})~~,
\end{eqnarray}
\begin{eqnarray}
  \label{f4}
  f_4\equiv q^{-{1\over 24}} \prod_{n=1}^\infty (1 - q^{2n -1})~~.
\end{eqnarray}
In the case of a real argument $q=e^{- \pi t}$ they transform as follows
under the modular transformation $t\rightarrow 1/t$:
\begin{eqnarray}
\label{mtf1}
f_1(e^{-\frac{\pi}{t}})=\sqrt{t}f_1(e^{-\pi t})~~~~,~~~~
f_2(e^{-\frac{\pi} {t}})=f_4(e^{-\pi t})~~~,~~~
f_3(e^{-\pi t})=f_3(e^{-\frac{\pi}{ t}})~.
\end{eqnarray}
The following relations are also useful:
\begin{eqnarray}
\Theta_{2,3,4} ( 0| it ) = f_1(e^{-\pi t}) \,f_{2,3,4}^{2}(e^{-\pi t})~~~;~~~\lim_{\nu \rightarrow 0}
\frac{\Theta_{1} (
  \nu | it )}{2 \sin \pi \nu} = f_{1}^{3}(e^{-\pi t})
\label{usere34}
\end{eqnarray}

\section{{\bf{Open String Channel}}}
\label{app2}
In this Appendix we  compute the one-loop vacuum
amplitude in the open string channel, by extending the techniques
developed in Ref.~\cite{BAPO}. We start by considering the action of an
open string in a background represented by an $SU(N)$ gauge field
$A_\mu^a$, being $\mu$ and $a$ respectively the Lorentz and gauge
indices. We choose only one end of the string to be charged under
the gauge field, say the one parameterized with $\sigma=0$, and,
for the sake of simplicity, we consider a gauge field pointing along a definite
direction in the gauge group, so that we can write
$A_{\nu}^a=\frac{1}{2}F_{\mu\nu}^a X^\mu$ where $F_{\mu \nu}$ is
taken constant. The string action  turns out to be:
\begin{eqnarray}
S=- \frac{1}{4\pi\alpha'} \int\! d^2\sigma \left[\eta^{\alpha \beta}
\partial_\alpha X^\mu \partial_\beta X_\mu - i \bar{\psi}^\mu \rho^\alpha
\partial_\alpha \psi_\mu\right]+\frac{1}{2}\int d\tau F_{\mu\nu}
\left[X^\nu \partial_\tau  X^\mu-\frac{i}{2}\bar{\psi}^\nu\rho^0
\psi^\mu\right]
\label{action}
\end{eqnarray}
Since the external field couples to the boundary, its effect is to modify
the boundary conditions of the string coordinates. In order to determine
such  modifications it is convenient to introduce the following sets of
coordinates:
\begin{eqnarray}
\label{cooz}
\!\!\!\!\!\!\! X_f^{\pm}\equiv\frac{X^0{\pm} X^1}{\sqrt{2}}
\,\,\,\,\,\,\,\,\,\,\,\,\,\,\,\,\,\,\,\,\,&
{\rm and}&X_g^{\pm}\equiv\frac{X^2{\pm} iX^3}{\sqrt{2}}~~;\\
\label{coopsi}
\psi_f^{\pm}\,_{R,L}\equiv\frac{\psi^0_{R,L}{\pm}\psi^1_{R,L}}{\sqrt{2}}
\,\,\,\,& {\rm and} &
\psi_g^{\pm}\,_{R,L}\equiv\frac{\psi^2_{R,L}{\pm} i\psi^3_{R,L}}{\sqrt{2}}~~,
\end{eqnarray}
where $\psi_R$ and $\psi_L$ are the two components with opposite chirality
of the
Majorana-Weyl world-sheet spinor $\psi$.

The surface terms, that arise in the variation of the action in
Eq. (\ref{action}) with a field strength $F$ given by Eq.
(\ref{effe}), vanish if the string
coordinates, previously introduced, satisfy the conditions:
\begin{eqnarray}
&&\label{bcond0}
\partial_\sigma X_f^{\pm}|_{\sigma=0}=\mp f\partial_\tau X_f^{\pm}|_{\sigma=0}~~,
\,\,\,\,\,\,\,\,\,\,\,\,\,\,\,\,\,\,\,\,\,\,
\partial_\sigma X_{g}^{\pm}|_{\sigma=0}=\mp i g\partial_\tau
X_{g}^{\pm}|_{\sigma=0}~~,
\\
&&\label{bcpsi0}
\psi_f^{\pm}\,_{R}|_{\sigma=0}=\frac{(1{\pm} f)}{(1\mp f)}\,
\psi_f^{\pm}\,_{L}|_{\sigma=0}~~,\,\,\,\,\,\,\,\,\,\,\,\,\,\,\,\,
\psi_g^{\pm}\,_{R}|_{\sigma=0}=\frac{(1{\pm} ig)}{ (1\mp ig)}
\psi_g^{\pm}\,_{L}|_{\sigma=0}~~
\end{eqnarray}
at $ \sigma=0$, together with the standard boundary conditions at
$\sigma=\pi$
\begin{equation}
\label{bcondpi}
\partial_\sigma X_{f}^{\pm}|_{\sigma=\pi}=0~,
\,\,\,\,\,\,\,\,\,\,\,\,\,\,\,
\partial_\sigma X_{g}^{\pm}|_{\sigma=\pi}=0~,
\end{equation}

\begin{equation}
\label{confe}
\psi_f^{\pm}\,_{R}|_{\sigma=\pi}+(-1)^a\psi_f^{\pm}\,_{L}|_{\sigma=\pi}=0~,\,\,\,\,\,\,\,\,\,\,\,\,\,\,\,\,
\psi_g^{\pm}\,_{R}|_{\sigma=\pi}+(-1)^a\psi_g^{\pm}\,_{L}|_{\sigma=\pi}=0~,
\end{equation}
with $a=0,1$ in the NS and R sector, respectively. The mode
expansions of the bosonic  coordinates in Eq. (\ref{cooz})
take the form
\begin{eqnarray}
X^{\pm}_f( \sigma,\tau)=x^{\pm}_f  +i\sqrt{2\alpha'} \sum_{n \in
Z}\frac{\alpha^{\pm}_{\,_f\,n} } {(n \pm i\epsilon_f) }
\phi_{\,_f\,n}^{\pm} (\sigma,\tau),&& \label{bosex}
\end{eqnarray}
where the oscillators modes are defined as follows:
\begin{eqnarray}
&&\alpha^\pm_{\,_f\,n}= \sqrt{ n{\pm} i\epsilon_f }\,a^\pm_n
\,\,\,\,n\geq 0\,\,\,\,\,
{\rm and}\,\,\,\,\,
\alpha^\pm_{\,_f\,-n }= \sqrt{n\mp i\epsilon_f
}\,a_{-n}^\pm\,\,\,\,\,\,n>0\\
&&\phi_{\,_f\,n}^{\pm}=e^{ -i \left(n{\pm} i\epsilon_f\right) \tau }
\cos\left[(n {\pm} i \epsilon_f)\sigma \mp i \pi\epsilon_f \right]~.
\label{wave}
\end{eqnarray}
For the coordinate $X_g$ we simply replace the index $f$ with $g$ and the
Fourier modes $a_n$ with $b_n$.
In the previous equations we have defined:
\begin{eqnarray}
\epsilon_f \equiv \frac{1}{\pi}{\rm arcth}f~,\,\,\,\,\,\,\,\,\,\,\,\,\,\,\,\,\,
\epsilon_g \equiv \frac{1}{\pi}{\rm arcth}(ig)
\label{eppar}
\end{eqnarray}
Analogously, the mode expansions of the  fermionic coordinates,
satisfying the conditions in Eqs. (\ref{bcpsi0}) and (\ref{confe}),
turn out to be:
\begin{eqnarray}
\label{mopsi}
\psi_f^{\pm}\,_{R,L}=\sqrt{2\alpha'}\!\!\sum_{n\in {\cal
    Z}+\frac{a+1}{2}}
d_{n}^{\pm}\chi_{\,_f\,n}^{\pm}\,_{R,L}(\sigma,\tau) \,
\,\,\,\,\,\,\,\,\,\,\,\,\,\,
\psi_g^{\pm}\,_{R,L}=\sqrt{2\alpha'}\!\!\sum_{n\in {\cal
    Z}+\frac{a+1}{2}}
h_{n}^{\pm}\chi_{\,_g\,n}^{\pm}\,_{R,L}(\sigma,\tau)
\end{eqnarray}
with
\begin{eqnarray}
\label{chipm}
\chi^{\pm}_{\,_f\,n}\,_R(\sigma,\tau)=\frac{1}{\sqrt 2}
e^{-i(n{\pm} i\epsilon_{f})(\tau-\sigma){\pm} \pi
\epsilon_{f}}, \,\,\,\,\,\,
\chi^{\pm}_{\,_f\,n}\,_L(\sigma,\tau)=\frac{1}{\sqrt 2}
e^{-i(n{\pm} i\epsilon_{f})(\tau+\sigma){\mp} \pi
\epsilon_{f}}~.
\end{eqnarray}
The Fourier
modes for the fields $\psi^{\pm}_g$ are, again, obtained by
replacing the index $f$ with $g$. Furthermore, it is useful in the
forthcoming discussion, to give also the relations between the
parameters defined in Eq. (\ref{nufnug2}) with the one given in
Eq. (\ref{eppar}):
\begin{eqnarray}
\epsilon_f=i\nu_f\qquad{\rm and}\qquad\epsilon_g=-i\nu_g.
\label{rne}
\end{eqnarray}
The  canonical quantization procedure leads to the following commutation
relations for the Fourier modes:
\begin{eqnarray}
&&[ x^+_f, x^-_f] =-i \frac{2\alpha' \pi}{f}  \hspace{1cm}
[a^{+}_{n},a^{-}_{-n}] =-1  \hspace{1cm}
\{ d_{n}^+,d_{m}^-\}=-\delta_{n+m}
\label{comm1} \\
&&  [ x^+_g, x^-_g]=
\frac{2\alpha' \pi}{g}
 \hspace{1.4cm} [b^{+}_{n},b^{-}_{-n}] =1\! \hspace{1.4cm} \{ h_{n}^+,h_{n}^-\}
=\delta_{n+m}~.
\label{comm}
\end{eqnarray}
Furthermore, with the help of Eqs. (\ref{bosex}) and
(\ref{mopsi}), one can compute the Virasoro generator $L_0$:
\begin{eqnarray}
L_0\!\!\!\!\!\!\!\!&&=L_{0}^\perp  - \sum_{n=-\infty}^{+\infty}:
\alpha^+_{\,_f\,n}{\cdot} \alpha^-_{\,_f\,-n}:+
\sum_{n=-\infty}^{+\infty}:\alpha^+_{\,_g\,{n}}{\cdot}
\alpha^-_{\,_g\,-n}:+\, c(a)\nonumber\\
&&-\!\!\sum_{n\in
\bf{Z}+\frac{a+1}{2}}\!\left(n+i\epsilon_f\right)\! :
d^-_{-n}{\cdot}d^+_{n}: +\!\!\sum_{n\in
\bf{Z}+\frac{a+1}{2}}\!\left(n+i\epsilon_g\right)\! :
h^-_{-n}{\cdot}h^+_{n}:\nonumber\\ &&\label{lm}
\end{eqnarray}
where $L_0^\perp$ denotes the standard contribution
coming from the direction orthogonal to the brane, and
the constant  $c(a)$ is the  zero point energy, whose value is  corrected
by  the presence of the
gauge field~\cite{BAPO}:
\begin{eqnarray}
&&c(0)=\frac{1}{2}i\epsilon_f(1-i\epsilon_f)+\frac{1}{2}i\epsilon_g(1-i
\epsilon_g)-
\frac{\epsilon^2_f}{2}-\frac{\epsilon^2_g}{2}\label{c0}-\frac{1}{2}~, \\
&&c(1)=0~.\label{c1}
\end{eqnarray}
Now we have all the ingredients to compute the one-loop free
energy given in Eq.~(\ref{Z}). It is the sum of
six terms:
\begin{eqnarray}
Z= Z_e^{NS} +  Z_e^{NS (-1)^F} +  Z_e^{R} +  Z_g^{NS} + Z_g^{NS(-1)^F}+  Z_g^{R(-1)^f}~.
\label{componenti}
\end{eqnarray}
In order to evaluate
each term,  we notice that the external field
causes essentially two modifications in the one-loop vacuum amplitude in Eq.
\eqref{Z} with respect to the case without gauge field. One  concerns
the oscillation frequencies of the longitudinal coordinates and the other
the zero mode contributions. In particular it is easy to see
that the oscillation frequencies get shifted by ${\pm} i\epsilon_f$ in the
$0,1$ plane and ${\pm} i\epsilon_g$ in the $(2,3)$ plane:
\begin{equation}
\begin{split}
\prod_{n=1}^\infty\left(\frac{1}{1-k^{2n}}\right)^8\longrightarrow &
\prod_{n=1}^\infty\left(\frac{1}{1-k^{2n}}\right)^4
\left(\frac{1}{1-k^{2n} e^{2\pi i\tau\epsilon_f}}\right)
\left(\frac{1}{1-k^{2n} e^{-2\pi i\tau\epsilon_f}}\right)\\
\times &\left(\frac{1}{1-k^{2n} e^{2\pi i\tau\epsilon_g}}\right)
\left(\frac{1}{1-k^{2n} e^{-2\pi i\tau\epsilon_g}}\right)
\end{split}
\end{equation}
with $k=e^{-\pi\tau}$. Analogous modifications occur in the
fermionic calculation. The contribution of the bosonic zero modes
to the partition function, instead, requires some care because of
the anomalous commutation relations satisfied by the coordinates
$x^{\pm}_{(f;g)}$, explicitly given in Eq. (\ref{comm1}).
Due to them, as explained
in Ref.~\cite{BAPO}, we have to compute the density of the quantum
states. By analogy with the case of  the conjugate variables
$[x,p]=i\frac{h}{2\pi}$, where such a density is simply given by
$\rho=1/h$, from Eqs. (\ref{comm1}) and (\ref{comm}), we deduce in our
case the following expression:
\begin{eqnarray}
\rho=-i\frac{f\,g}{(4\pi^2\alpha')^2}~~.
\label{den}
\end{eqnarray}
The contribution of the bosonic zero modes to the free energy is
given by:
\begin{eqnarray}
\!\!{\rm
Tr}\left[e^{-2\pi\tau\left(\frac{y^2}{(2\pi)^2\alpha'}-i\epsilon_f
a_{0}^-a_{0}^++i\epsilon_g b_{0}^-b_{0}^+\right)}\right] =\!
e^{-\frac{y^2\tau}{2\pi\alpha'}}\left[\frac{1}{1- e^{-2\pi
i\tau\epsilon_f}}\right]
\left[\frac{1}{1-e^{-2\pi i\tau\epsilon_g}}\right]~~.\nonumber
\label{zm}
\end{eqnarray}
Let us consider the fermionic zero mode contribution to the free
energy arising from the Ramond sector. It is well known that it
is divergent and must be
regularized~\cite{BDFLPRS}. It is more convenient to perform the regularization in the Euclidean space. At this aim we
first introduce the following operators:
\begin{eqnarray}
\psi_0^0=\frac{d^+_0+d_0^-}{\sqrt{2}}~, \qquad
\psi_0^1=\frac{d^+_0-d_0^-}{\sqrt{2}}~, \qquad
\psi_0^2=\frac{h^+_0+h_0^-}{\sqrt{2}}~, \qquad
\psi_0^3=\frac{h^+_0-h_0^-}{\sqrt{2}i}~
\label{zm1}
\end{eqnarray}
that, together with the zero modes associated to the transverse
directions $\psi_0^i$ ($j=4, \dots 9$), satisfy the usual
anticommutation rules $\{\psi_0^\mu\psi_0^\nu\}=\eta^{\mu\nu}$.
Then we perform a Wick rotation $\psi_0^0=i\psi_0^{10}$.
Furthermore, it is convenient to introduce the raising and
lowering operators defined as follows:
\begin{eqnarray}
e_1^{\pm}=\pm d_0^\pm\,\,\,\,;\,\,\,\,e_2^{\pm}=h_0^\pm
\qquad
e^{\pm}_j=\frac{\psi^{2j-2}_0{\pm}i\psi_0^{2j-1}}{\sqrt{2}}
\,\,\,{\rm with} \,\,j=3\dots 5
\label{eop}
\end{eqnarray}
satisfying the algebra:
\begin{eqnarray}
\{e^+_{\rm \bf a},e^-_{\rm \bf b}\}=\delta_{{\rm \bf a},{\rm \bf b}}~.
\label{eaeb}
\end{eqnarray}
The Hilbert space, associated to each couple of  operators in Eq.
(\ref{eop}), is two-dimensional and it is spanned  by the states
$|s_{\rm \bf a}={\pm}1\rangle$, being $s$ the eigenvalues of the
number operators
\begin{eqnarray}
N_{\rm \bf a}=[e^+_{\rm \bf a},e_{\rm \bf a}^-] \label{noper}
\end{eqnarray}
In terms of these latter, the fermionic zero modes in the R-sector
become
\begin{eqnarray}
d_0^-d_0^+=\frac{N_1-1}{2},\qquad h_0^-h_0^+=\frac{1-N_2}{2}
\end{eqnarray}
and the GSO, projector together with the orbifold action, takes the
following form~\cite{REV,BCR0011}:
\begin{eqnarray}
(-1)^{F_0}=\prod_{k=1}^5 N_k,\qquad g=e^{i\frac{\pi}{2}(N_4+ N_5)}=-N_4N_5.
\label{fg}
\end{eqnarray}
Let us  compute explicitly the traces over the fermionic zero
modes appearing in Eq. (\ref{Z}). In particular we have:
\begin{eqnarray}
\label{traccia1}
{\rm Tr}_{\,R}\left(e^{-2\pi\tau(-i\epsilon_f d_0^-d_0^++i\epsilon_g
h_0^-h_0^+)}(-1)^{G_{\beta\gamma}}\right)
=4\left(1+e^{-i\pi\tau\epsilon_f}\right)\left(1+e^{-i\pi\tau\epsilon_g}\right)
\end{eqnarray}
and
\begin{eqnarray}
\label{traccia2} &&{\rm Tr}_{\,R}\left(e^{-2\pi\tau(-i\epsilon_f
d_0^-d_0^++i\epsilon_g h_0^-h_0^+)} g(-1)^{F_0}\right)\nonumber\\
&&=-\lim_{x\rightarrow 1}{\rm Tr}_{\,R}\left(e^{i\pi\tau
[\epsilon_f(N_1-1)-\epsilon_g(1-N_2)]}\prod_{k=1}^5x^{N_k}N_1N_2N_3
\right){\rm Tr}\left(x^{-2\gamma_0\beta_0}\right)
\end{eqnarray}
where we have introduced the regulator
$x^{\sum_kN_k}x^{-2\gamma_0\beta_0}$ in order to have a finite
result.
Therefore by using the following relations:
\begin{eqnarray}
{\rm
Tr}\left(x^{N_k}N_k\right)=\left(x-\frac{1}{x}\right),\qquad{\rm
Tr}\left(x^{N_k}\right)=\left(x+\frac{1}{x}\right)\,\,\,\,\,\,
\end{eqnarray}
\begin{eqnarray}
{\rm Tr}\left(e^{i\pi\tau\epsilon_f
N_k}x^{N_k}N_k\right)=\left(xe^{i\pi\tau\epsilon_f}-\frac{1}{xe^{i\pi\tau\epsilon_f}}\right),
\,\,\,\,\,\,\,\,\,\,\,\,\,\,\,\,
 {\rm Tr}\left(x^{-2\gamma_0\beta_0}\right)=\frac{1}{1-x^2} ,
\end{eqnarray}
we get that the zero mode contribution of the $R(-1)^F$ sector to $Z_g$ is
\begin{eqnarray}
{\rm Tr}_{\,R}\left(e^{-2\pi\tau(-i\epsilon_f
d_0^-d_0^++i\epsilon_g h_0^-h_0^+)}g(-1)^{F_0}\right)= -16 e^{-\pi
i\tau \epsilon_f} e^{-i\pi \tau \epsilon_g}\sin{\pi
\tau\epsilon_f}\sin{\pi \tau\epsilon_g}.
\end{eqnarray}
Finally, by collecting all the results, and observing that in
Eq. (\ref{den}) one can write
\begin{eqnarray}
i{f\,g}={\sin{\pi\nu_f\sin\pi\nu_g} \sqrt{-{\rm det}(\eta +\hat
F)}}
\label{den1}
\end{eqnarray}
we get the following expressions for the various terms
defined in Eq. (\ref{componenti}):
\begin{eqnarray}
\label{zeta} Z&=&-N\int d^4x \sqrt{-{\rm det}
\left(\eta+\hat{F}\right)}\left[\frac{\sin \pi \nu_f \sin\pi
\nu_g}{(4\pi^2\alpha')^2}\right] \tilde{Z}\nonumber
\end{eqnarray}
with
\begin{eqnarray}
\tilde{Z}_e^{NS}&&=\frac{1}{4}\, \int_{0}^\infty
\frac{d\tau}{\tau}e^{-\frac{y^2\tau}{2\pi\alpha'}} \frac{f_3^4(k)
\Theta_{3}(i\nu_f\tau|i\tau)\Theta_{3}(i\nu_g\tau|i\tau)}{f_1^4(k)
\Theta_{1}(i\nu_f\tau|i\tau)\Theta_{1}(i\nu_g\tau|i\tau)},
\\
 \tilde{Z}_e^{NS(-1)^F}&&=\,\,-\frac{1}{4}
\int_{0}^\infty \frac{d\tau}{\tau}e^{-\frac{y^2\tau
}{2\pi\alpha'}} \frac{f_4^4(k)
\Theta_{4}(i\nu_f\tau|i\tau)\Theta_{4}(i\nu_g\tau|i\tau)}{
f_1^4(k)\Theta_{1}(i\nu_f\tau|i\tau)\Theta_{1}(i\nu_g\tau|i\tau)},\\
 \tilde{Z}_e^{R}&&= \,\,-\frac{1}{4}
\int_{0}^\infty \frac{d\tau}{\tau}e^{-\frac{y^2\tau
}{2\pi\alpha'}} \frac{f_2^4(k)
\Theta_{2}(i\nu_f\tau|i\tau)\Theta_{2}(i\nu_g\tau|i\tau)}{f_1^4(k)
\Theta_{1}(i\nu_f\tau|i\tau)\Theta_{1}(i\nu_g\tau|i\tau)},
\label{zetar}\\
\tilde{Z}_h^{NS}&&=\,\,
\int_{0}^\infty \frac{d\tau}{\tau}e^{-\frac{y^2 \tau}{2\pi\alpha'}}
\frac{\Theta_{4}^2(0|i\tau)\Theta_{3}(i\nu_f\tau|i\tau)
\Theta_{3}(i\nu_g\tau|i\tau)}
{\Theta_{2}^2(0|i\tau)
\Theta_{1}(i\nu_f\tau|i\tau)\Theta_{1}(i\nu_g\tau|i\tau)},\label{zgns}\\
\tilde{Z}_h^{NS(-1)^F}&&=\,\,\,\,-
\int_{0}^\infty \frac{d\tau}{\tau}e^{-\frac{y^2 \tau}{2\pi\alpha'}}
\frac{\Theta_{3}^2(0|i\tau) \Theta_{4}(i\nu_f\tau|i\tau)\Theta_{4}(i\nu_g\tau|i\tau)}
{\Theta_{2}^2(0|i\tau)
\Theta_{1}(i\nu_f\tau|i\tau)\Theta_{1}(i\nu_g\tau|i\tau)},\label{zgnsf}
\end{eqnarray}
while
\begin{eqnarray}
{Z}_h^{R(-1)^F}&&=  iN\int d^4 x \frac{fg}{(4\pi^2\alpha')^2}
\int_{0}^\infty \frac{d\tau}{\tau}e^{-\frac{y^2
\tau}{2\pi\alpha'}}~~.
 \label{zetateta}
\end{eqnarray}
The last contribution can be written in terms of
$F_{\alpha\beta}^a
\tilde{F}^{a\,\alpha\beta}$, with
$\tilde{F}^{a}_{\alpha\beta}=\frac{1}{2}\epsilon_{\alpha\beta\gamma\delta}
F^{a\,\gamma\delta}$, using the relation
\begin{equation}
f\, g = \frac{(2\pi\alpha')^2}{8}F_{\alpha\beta}^a
\tilde{F}^{a\,\alpha\beta},
\label{fg56}
\end{equation}
giving the last term in Eq. (\ref{zetatet}).

In the final part of this Appendix we expand Eq. (\ref{zetatet}) up to
quadratic terms in the gauge field. We need the following expansions:
\begin{equation}
\Theta_3 ( i \nu_f \tau | i \tau ) = \Theta_3 ( 0 | i \tau ) \left[ 1
  - 4 f^2 \tau^2 \sum_{n=1}^{\infty} \frac{k^{2n-1}}{(1 + k^{2n-1})^2}
  + \dots \right]~,
\label{exp331}
\end{equation}
\begin{equation}
\Theta_4 ( i \nu_f \tau | i \tau ) = \Theta_4 ( 0 | i \tau ) \left[ 1
  + 4 f^2 \tau^2 \sum_{n=1}^{\infty} \frac{k^{2n-1}}{(1 - k^{2n-1})^2}
  + \dots \right]
\label{exp44}
\end{equation}
and
\begin{equation}
\lim_{\nu \rightarrow 0} \frac{2 \sin \pi \nu}{\Theta_1 ( i \nu \tau|
  i \tau)} = - \frac{1}{i \tau f_{1}^{3} (k)}~.
\label{exp11}
\end{equation}
Inserting them in Eq. (\ref{zetatet}), reintroducing the field
strength of the gauge
field by means of the following
equation~\footnote{The generators of $SU(N)$ are normalized as:
${\rm Tr}[T^a\,T^b]=\frac{1}{2}\delta^{ab}$.}:
\begin{eqnarray}
f^2 -g^2 &=&- \frac{(2 \pi \alpha')^2}{4} F_{\alpha \beta}^{a}
F^{a \alpha \beta}
\label{f297}
\end{eqnarray}
and using Eq. (\ref{fg56}), we get for the term quadratic in the gauge
field the following expression:
\begin{eqnarray}
Z_h^o\left(F^2\right)\!\!&=&\!-\,\frac{N}{ 8 \pi^2 }
\int d^4 x
\left[ - \frac{1}{4} F_{\alpha \beta}^{a} F^{a\alpha \beta } \right]
\int_{0}^{\infty} \frac{d
\tau}{\tau}
{e}^{-\frac{y^2 \tau}{2 \pi \alpha' } } \nonumber\\
&&\!{\times} \,\frac{1}{2} \prod_{n=1}^{\infty}\frac{ [1 +  k^{ 2n-1} ]^4
[1 - k^{2n-1} ]^4}{[1 -  k^{ 2n} ]^4
[1 + k^{2n} ]^4 } \sum_{n=1}^{\infty} \left[
\frac{k^{2n-2}}{( 1 + k^{2n-1})^2 } +
\frac{k^{2n-2}}{( 1 - k^{2n-1})^2 }    \right]\nonumber\\
&& \! + \, \frac{\, iN }{ 32 \pi^2  } \int d^4 x  F_{\alpha\beta}^a
\tilde{F}^{a\,\alpha\beta}
\int _0^\infty \frac{d\tau}{\tau}
e^{-\frac{y^2\,\tau}{2\pi\alpha'}}~. \label{f2term9}
\end{eqnarray}
Notice that in the previous calculation we do not need to compute the
quadratic term in $f$ and $g$ coming from the piece in front of the
square bracket in Eq. (\ref{zetatet}) because its coefficient is zero.
Finally, by using the following identities, which can be proven with the
help of the heat equation (\ref{iden})
\begin{eqnarray}
&&\sum_{n=1}^\infty \frac{k^{2n-1}}{(1-k^{2n-1})^2} = -k \frac{d}{dk}\left[
\frac{1}{2} \log \prod_{n=1}^\infty (1-k^{2n})+
\log \prod_{n=1}^\infty (1-k^{2n-1})\right]~,
\nonumber\\
&&\sum_{n=1}^\infty \frac{k^{2n-1}}{(1+k^{2n-1})^2} = k
\frac{d}{dk}\left[\frac{1}{2}
\log \prod_{n=1}^\infty (1-k^{2n})+\log \prod_{n=1}^\infty
(1+k^{2n-1})\right]~,\nonumber\\
&&\sum_{n=1}^\infty \frac{k^{2n}}{(1+k^{2n})^2}=  k
\frac{d}{dk}\left[\frac{1}{2}
\log \prod_{n=1}^\infty (1-k^{2n})+\log \prod_{n=1}^\infty
(1+k^{2n})\right]~,\nonumber\\
&&
\label{rel}
\end{eqnarray}
we get the following compact expression for  Eq. (\ref{f2term9}):
\begin{eqnarray}
Z_g^o(F^2)\!\!&=&  \!-\frac{N}{ 8 \pi^2 } \int d^4 x
\left[-\frac{1}{4} F_{\alpha \beta}^{a} F^{a \alpha \beta}\right]
\int_{0}^{\infty} \frac{d \tau}{\tau}
{e}^{-\frac{y^2 \tau}{2 \pi \alpha' } }
\left[\frac{f_3(k)f_4(k) }{f_1(k)
f_2( k)}\right]^4
2 k \frac{d}{dk} \log \left[\frac{f_3(  k)}{f_4( k) }\right]\nonumber\\
&&\! +\frac{i N}{32\pi^2} \int d^4 x F_{\alpha\beta}^a
\tilde{F}^{a\,\alpha\beta} \int _0^\infty \frac{d\tau}{\tau}
e^{-\frac{y^2\,\tau}{2\pi\alpha'}}, \label{g1}
\end{eqnarray}
which reproduces Eq. (\ref{gau56}).

\section{{\bf{Closed String Channel}}}
\label{app4}
In this Appendix we derive Eqs. (\ref{op2}) and
(\ref{closed}), which added together provide the interaction amplitude
between a stack of $N$ D$3$ branes and a brane dressed with an
external field. The two contributions correspond  to the
propagation of  untwisted and twisted  closed string
states. They were computed in the Appendix
of Ref.~\cite{d3d7} in the case of vanishing external gauge field.

The boundary state describing a $Dp$-brane without any gauge field on
its world-volume and
living on the orbifold $\mathbb{C}^2/\mathbb{Z}_2$ is given in
Ref.~\cite{d3d7} and
it is the sum of two terms, one relative to the
untwisted sector and
the other to the twisted one.
The boundary state
describing a dressed brane completely transverse to the orbifold
is given, for the untwisted sector U, by :
\begin{eqnarray}
\label{bound1}
\ket{{\rm D}p;F}^U =\,\, \frac{T_p}{2\sqrt{2}}\, \left(\,
\ket{{\rm D}p;F}_{\rm NS}^U \,+ \,\ket{{\rm D}p;F}_{\rm R}^U\, \right)
\end{eqnarray}
where $\ket{{\rm D}p;F}_{\rm NS}^U$ and $\ket{{\rm D}p;F}_{\rm R}^U$
are the usual boundary states for a dressed bulk D$p$-brane
given in Refs.~\cite{DL99121,DLII}.

In the twisted sector T, instead, the dressed boundary state is given by:
\begin{eqnarray}
\label{boundt}
\ket{{\rm D}p;F}^T \, = \,- \,\,
\frac{T_p}{2\sqrt{2}\,\pi^2 \alpha' } \,
\left(\,\ket{{\rm D}p;F}_{{\rm NS}}^T + \ket{{\rm D}p;F}_{{\rm R} }^T\, \right)
\end{eqnarray}
where
\begin{eqnarray}
\label{proi}
\ket{{\rm D}p;F}_{{\rm NS,R}}^T
= \frac{1}{2}\,\left(\,\ket{{\rm D}p;F\, +}_{{\rm NS,R} }^T
\,+\, \ket{{\rm D}p;F \, -}_{{\rm NS,R} }^T\, \right)~~,
\end{eqnarray}
and the Ishibashi states  $\ket{{\rm D}p;F \, \eta=\pm}_{\rm NS,R}^T$ are
\begin{eqnarray}
\label{bound2}
\ket{{\rm D}p;F \,\eta}^T_{\rm NS,R}=
\ket{{\rm D}p_X;F}^T\ket{{\rm D}p_\psi; F \,\eta}_{\rm NS,R}^T
\end{eqnarray}
with\footnote{In Eq. (\ref{bound2})
we omit the ghost and superghost contributions
which are not affected by the
orbifold projection.}
\[
|{\rm D}p_X;F \rangle^T  =\sqrt{{-\rm det(\eta + \hat F)}} \, \delta^{(5-p)}
({\widehat q}^i-y^i)
\]
\[
 \times \prod_{n=1}^{\infty} \left[ {\rm e}^{-\frac{1}{n}
\alpha_{-n}^\alpha
M_{\alpha \beta}\tilde\alpha_{-n}^\beta}
{\rm e}^{\frac{1}{n}
\alpha_{-n}^i\tilde\alpha_{-n}^i} \right]
\times\,\prod_{r=\frac{1}{2}}^{\infty} {\rm e}^{- \frac{1}{r}
\alpha_{-r}^\ell\tilde\alpha_{-r}^{\,\ell}}
\prod_{\beta}
|p_\beta = 0\rangle
\prod_{{i}}
|p_i = 0\rangle~~,
\]
\[
|{\rm D}p_{\psi} ;F \, \eta \rangle_{NS}^T  =
\prod_{r=\frac{1}{2}}^{\infty} \left[
{\rm e}^{{\rm i}\eta\psi_{-r}^\alpha M_{\alpha \beta}
\tilde \psi_{-r}^\beta}
 {\rm e}^{-{\rm i}\eta\psi_{-r}^i
\tilde \psi_{-r}^i} \right]
\prod_{n=1}^{\infty} e^{{\rm i}\eta\psi_{-n}^\ell
\tilde \psi_{-n}^{\,\ell}} |{\rm D}p_{\psi} ,
\eta \rangle ^{(0)\,\,T}_{\rm NS}~~,
\]
\begin{equation}
|{\rm D}p_{\psi} ;F \, \eta \rangle_{R}^T = \frac{1}{\sqrt{{-\rm det(\eta + \hat F)}}}
\prod_{n=1}^{\infty} \left[ {\rm e}^{{\rm i}\eta\psi_{-n}^\alpha
M_{\alpha \beta}
\tilde \psi_{-n}^\beta}
{\rm e}^{-{\rm i}\eta\psi_{-n}^i \tilde \psi_{-n}^i} \right]
\prod_{r=\frac{1}{2}}^{\infty} {\rm e}^{{\rm i}\eta\psi_{-r}^\ell
\tilde \psi_{-r}^{\,\ell}} |{\rm D}p_{\psi} ; F \,
\eta \rangle ^{(0)\,\,T}_{\rm R}.
\end{equation}
In these expressions the longitudinal indices $\alpha, \beta$ take
the values $0,1, \ldots p$,
the transverse index $i$ takes the values  $p+1,
\ldots, 5$, while the index $\ell$ labels the orbifold directions.
Furthermore the matrix $M$ is defined by:
\begin{eqnarray}
M^{\alpha}_{\beta} = \left[ (1-\hat{F}) (1+\hat{F})^{-1}
\right]^{\alpha}_{\beta} .
\end{eqnarray}
The zero-mode part of the boundary state in the NS-NS sector has the
same structure as the one without gauge field, while in the R-R sector
it reads as:
\begin{eqnarray}
|{\rm D}p_{\psi} ; F \,
\eta \rangle ^{(0)\,\,T}_{\rm R} =
\left(C\gamma^0...\gamma^p\frac{1+i\eta_1\tilde{\gamma}}{1+i\eta_1};
e^{-\frac{1}{2}\hat F_{\alpha\beta}\gamma^{\alpha}\gamma^{\beta}};
\right)_{ab}|a\rangle |\widetilde b\rangle|D_\sgh,\eta_1\rangle_\R^{(0)}
\end{eqnarray}
where the symbol $; \,$  $\, ;$ means that we have to expand the
exponential
and then to antisymmetrize the indices of the $\gamma$ matrices.
The superghost zero mode contribution is unchanged with
respect to the untwisted sector and can be found
in Ref.~\cite{DL99121}. Here the $\gamma$ matrices reproduce the
Clifford algebra in six dimensions and $\tilde{\gamma}\equiv
\prod_{i=0}^5\gamma^i$.

We have now introduced all the ingredients necessary to compute the
tree level closed string amplitude given
in Eqs. (\ref{ze32}) and (\ref{zg32}), that are relative to the case $p=3$
we are interested in. The explicit expression for the
untwisted component $Z_{e}^c$ is:
\begin{eqnarray}
Z_{e}^c \!\!\! &&=   \frac{1}{4} \frac{1}{(8 \pi^{2} \alpha')^{2} }\int
d^4 x \sqrt{ - \mbox{det} ( \eta + \hat{F} )}
\int_{0}^{+\infty}
\frac{dt}{t^{3}} \frac{e^{- \frac{y^2 }{ 2 \pi \alpha' t} }}
{\prod_{n=1}^{+\infty}\mbox{det}
\left(1-M^{T} q^{2n}\right)\left( 1- q^{2n} \right)^{4} }
\nonumber\\
&&\times \left\{
\frac{1}{q}
\left[\prod_{n=1}^{+\infty}
{\mbox{det}(1 + M^{T} q^{2n-1})\left( 1+ q^{2n-1}\right)^{4}}
-\prod_{n=1}^{+\infty}{\mbox{det}\left( 1-M^{T} q^{2n-1} \right)
\left( 1- q^{2n-1}\right)^{4}}
\right] \right.
\nonumber\\
&&
\left.
-\frac{2^4}{\sqrt{ - \mbox{det} (\eta + \hat{F} )}}
 \prod_{n=1}^{+\infty}{ \mbox{det} (1 + M^{T} q^{2n} ) \left( 1+ q^{2n}
\right)^{4} }
\right\},
\label{untwisted}
\end{eqnarray}
where $q=e^{-\pi t}$.
We are not going to derive in detail  Eq. (\ref{untwisted}), but
we want to point out that it can be easily obtained from Eq. (A.2)
of Ref.~\cite{d3d7}, rewritten in the closed string
channel by adding in the NS-NS sector the Born-Infeld action and the
contribution of the gauge fields in the brane world-volume directions. Furthermore we need to make the
following substitutions:
\begin{eqnarray}
\prod_{n=1}^{\infty} ( 1  \pm q^{2n-1})^{4} \rightarrow
\prod_{n=1}^{\infty} \det ( 1 \pm M^T  q^{2n-1})
\label{sub45}
\end{eqnarray}
and
\begin{eqnarray}
\prod_{n=1}^{\infty} ( 1  \pm q^{2n} )^{4} \rightarrow
\prod_{n=1}^{\infty} \det ( 1 \pm M^T  q^{2n} )
 .
\label{sub46}
\end{eqnarray}
Analogous substitutions have to be done in the twisted sector. In
particular,
in the NS-NS sector the expression for the tree level closed string amplitude
given in Eq. (\ref{zg32}) is equal to:
\begin{eqnarray}
Z_h^{c ({\rm NS-NS})} \!\!\! & = &\frac{4} {(8 \pi^{2} \alpha')^{2}} \int d^4x
\sqrt{ - \mbox{det} ( \eta + \hat{F} )}
\int_{0}^{+\infty}
\frac{dt}{t}
e^{- \frac{y^2 }{ 2 \pi \alpha' t} }\nonumber\\
&\times &
\prod_{n=1}^{+\infty}
\frac{ \mbox{det} (1 + M^{T} q^{2n-1} )
\left( 1+ q^{2n}
\right)^{4} }
{\mbox{det} (1 - M^{T} q^{2n} )(1-q^{2n-1})^{4} }~~. \label{twisted0}
\end{eqnarray}
The zero modes in the NS-NS twisted sector coincide with those with vanishing gauge field. Hence, as shown in Ref. \cite{d3d7}, the spin structure
NS-NS$(-1)^{F}$ does not
contribute to the interaction in this sector.

In the  R-R sector we can proceed in an analogous way, but in this
case we have also a contribution from the zero modes that is divergent and
requires to be treated more carefully through
a suitable regularization. Let us evaluate it explicitly.
According to Ref.~\cite{BDFLPRS} we insert
the regulator ${\cal R}(x) = x^{2(F_0+G_0)}$ as follows:
\[
{}_\R^{T\,(0)}\!\langle {\rm D}3_{\psi},\eta_2 |
{\rm D}3_{\psi}; F \, \eta_1\rangle_\R^{(0)} \equiv
\lim_{x\to 1} {}_\R^{T\,(0)}\!\langle {\rm D} 3_{\psi},\eta_2 |
 \, {\cal R}(x) \,
|{\rm D}3_{\psi} ; F \,
\eta_1 \rangle ^{(0)\,\,T}_{\rm R}
\]
\begin{eqnarray}
=\lim_{x\to 1} \left[~{}_\R^{T\,(0)}\!\langle {\rm D}3_{\psi},\eta_2 |
 \, x^{2F_0} \,
|{\rm D}3_{\psi} ; F \,
\eta_1 \rangle ^{(0)\,\,T}_{\rm R}
\times~{}_\R^{(0)}\!\langle {\rm D}_\sgh,\eta_2 | \, x^{2G_0}
\, |{\rm D}_\sgh,\eta_1\rangle_\R^{(0)}~
\right]~~.
\label{prodfact}
\end{eqnarray}
where $(-1)^{F_0}=i\tilde{\gamma}$ and $G_0=-\gamma_0\beta_0$.

For the superghost part the regularization scheme leads to the same
result as in the untwisted sector~\cite{BDFLPRS,DL99121}:
\begin{eqnarray}
{}_\R^{(0)}\!\langle B_\sgh,\eta_2 | \, x^{2G_0} \,
|B_\sgh,\eta_1\rangle_\R^{(0)} &=&
\bra{-{3/2},-{1/2}}
{\rm  e}^{-i\eta_2\beta_0\widetilde\gamma_0}\,x^{-2\gamma_0\beta_0}\,
  {\rm e}^{i\eta_1\gamma_0\widetilde\beta_0}\ket{-{1/2},-{3/2}}
\nonumber \\
&=&\frac{1}{1 + \eta_1 \eta_2 x^2}~~.
\label{bs50}
\end{eqnarray}
In order to evaluate the matter part we introduce
the
operators
\begin{eqnarray}
\label{enne}
N_1\equiv \gamma^0\gamma^1\,\,\,\,\,;\,\,\,\,\,N_2\equiv
-i\gamma^2\gamma^3\,\,
\,\,\,;\,\,\,\,\,
N_3\equiv -i\gamma^4\gamma^5
\end{eqnarray}
and write
\begin{eqnarray}
(-1)^{F_0} = i {\tilde \gamma} = -i\,\prod_{k=1}^{3}
N_k=\prod_{k=1}^{3} \exp\left({i\,N_k\,{\pi}/{2}}\right)=
(-1)^{\frac{1}{2}(N_1+N_2+N_3)},
\label{f00}
\end{eqnarray}
where we have used that $\exp\left({i\,N_k\,{\pi}/{2}}\right) = i N_k$.
From the previous equation we can read $F_{\rm 0}=
\frac{1}{2} \sum_{k=1}^{3}N_k$
and thus the regulator for the fermionic zero-modes can be written as follows:
\begin{eqnarray}
\label{bs45b}
x^{2F_{\rm 0}} = x^{~\sum\limits_{k=1}^3 N_{k} }~~.
\end{eqnarray}
Substituting this expression in the matter part of
Eq. (\ref{prodfact}) we get:
\[
{}_\R^{T\,(0)}\!\langle {\rm D}3_{\psi},\eta_2 | \, x^{2 F_0} \,
|{\rm D}3_{\psi}; F \, \eta_1\rangle_\R^{(0)\,T}  = -
\delta_{\eta_1\eta_2;1}Tr[x^{2 F_0}]
\]
\begin{eqnarray}
+\delta_{\eta_1\eta_2;-1}
(\pi\alpha')^2F_{\alpha\beta} \tilde F^{\alpha\beta}\,\,
{\rm Tr}[x^{2 F_0}\tilde{\gamma}\gamma^0...\gamma^3]
\label{bpsibpsi}
\end{eqnarray}
where we have only kept those terms which, added to the ghost contribution in Eq. (\ref{bs50}), yields a non-zero result.
The traces appearing in the previous equation are easily evaluated
\begin{eqnarray}
{\rm Tr}[x^{2F_0}] & = &  \prod_{k=1}^{3} {\rm Tr}[x^{N_k}]=
\left(x+\frac{1}{x}\right)^3,\\
{\rm Tr}[x^{2F_0}\gamma^0\gamma^1\gamma^2\gamma^3] & = &
i {\rm Tr} [x^{N_1}N_1] {\rm Tr} [x^{N_2}N_2] {\rm Tr} [x^{N_3}]=
i\left(x-\frac{1}{x}\right)^2\left(x+\frac{1}{x}\right),\\
{\rm Tr}[x^{2F_0}\gamma^0\gamma^1\gamma^2\gamma^3\tilde{\gamma}] & = &
-i {\rm Tr} [x^{N_1}] {\rm Tr} [x^{N_2}] {\rm Tr} [x^{N_3}N_3]=
-i\left(x+\frac{1}{x}\right)^2\left(x-\frac{1}{x}\right).
\label{tracce}
\end{eqnarray}
By plugging them in Eq. (\ref{bpsibpsi}), including also
the ghost contribution and performing the $x\rightarrow 1$ limit
one gets:
\begin{eqnarray}
\label{fampie}
{}_\R^{T \, (0)}\!\langle {\rm D}3_\psi,\eta_2 |
{\rm D}3_{\psi}; F \, \eta_1\rangle_\R^{(0) \, T}= -4\delta_{\eta_1\eta_2;1} +
4i\delta_{\eta_1\eta_2;-1}
\left(\frac{(\pi\alpha')^2}{2}F^a_{\alpha\beta}\tilde F^{a\alpha\beta}\right)
\end{eqnarray}
where an extra factor $1/2$ has been introduced to take in account
the trace over the $SU(N)$ generators.
By adding to the previous expression the contribution of the non zero
modes it is
straightforward to write down the complete expression for the interaction in
the twisted R-R sector:
\begin{eqnarray}
&&Z_h^{c({\rm R-R})} \!\!\! = - \frac{4} {(8 \pi^{2} \alpha')^{2}} \int d^4x
\sqrt{ - \mbox{det} ( \eta + \hat{F} )}
\int_{0}^{+\infty}
\frac{dt}{t}
\frac{e^{- \frac{y^2 }{ 2 \pi \alpha' t} }}
{\prod_{n=1}^{+\infty}\mbox{det} (1 - M^{T} q^{2n} )(1-q^{2n-1})^{4} }
\nonumber\\
&&\times
\frac{\prod_{n=1}^{+\infty} {
 \mbox{det} \left( 1+M^{T} q^{2n} \right)}
{\left( 1+ q^{2n-1}
\right)^{4} }}{\sqrt{ - \mbox{det} ( \eta + \hat{F} )}}
+\frac{iN}{32\pi^{2}}\int d^{4}x F_{\alpha\beta}^a\tilde F^{a\alpha\beta}
\int_0^\infty\frac{dt}{t}
e^{- \frac{y^2 }{ 2 \pi \alpha' t} } .
\label{twisted}
\end{eqnarray}
The determinants of the terms containing the external gauge field
present in Eqs. (\ref{twisted0}) and (\ref{twisted}) can
be computed using the parametrization for $F$ given in Eq. (\ref{effe})
and one gets the following expressions in terms of the $\Theta$
functions defined in Appendix~\ref{app0}:
\begin{equation}
\prod_{n=1}^{\infty} \det \left(1 + M^T q^{2n-1}  \right)=
\Theta_3 ( \nu_f | it )  \Theta_3 ( \nu_g | it)
\prod_{n=1}^{\infty} ( 1 - q^{2 n})^{-2}~,
\label{prod1}
\end{equation}

\begin{equation}
\prod_{n=1}^{\infty} \det \left( 1 - M^T q^{2  n } \right)=
{e}^{\pi t/2} \frac{\Theta_1 ( \nu_f | it )}{2 \sin \pi \nu_f}
\frac{\Theta_1 ( \nu_g | it )}{2 \sin \pi \nu_g}
\prod_{n=1}^{\infty} ( 1 - q^{2 n})^{-2}~,
\label{prod2}
\end{equation}

\begin{equation}
\prod_{n=1}^{\infty} \det \left(1 + M^T q^{ 2  n } \right)=
{e}^{\pi t/2} \frac{\Theta_2 ( \nu_f | it )}{2 \cos \pi \nu_f}
\frac{\Theta_2 ( \nu_g | it )}{2 \cos \pi \nu_g}
\prod_{n=1}^{\infty} ( 1 - q^{2 n})^{-2}~,
\label{prod3}
\end{equation}

\begin{equation}
\prod_{n=1}^{\infty} \det \left(1 - M^T q^{ 2n-1} \right)=
\Theta_4 ( \nu_f | it )  \Theta_4 ( \nu_g | it)
\prod_{n=1}^{\infty} ( 1 - q^{2 n})^{-2}~.
\label{prod4}
\end{equation}
Inserting these formulas in Eqs. (\ref{untwisted}), (\ref{twisted0}),
(\ref{twisted}) and
using the definition of the functions $f_i$ given in Appendix A,
one easily gets Eqs. (\ref{op2}) and (\ref{closed}).

In the last part of this Appendix we expand Eq. (\ref{closed}) keeping
up to terms quadratic in the gauge field.
This can be done by using the following expansions:
\begin{equation}
\Theta_3 ( \nu_f | i t) = \Theta_3 ( 0 | i t) \left[1 + 4 f^2
  \sum_{n=1}^{\infty} \frac{q^{2n-1}}{(1 + q^{2n-1})^2} + \dots \right]~,
\label{exp33}
\end{equation}
\begin{equation}
\Theta_2 ( \nu_f | i t) = \Theta_2 ( 0 | i t) \left[1 + 4 f^2
  \sum_{n=1}^{\infty} \frac{q^{2n}}{(1 + q^{2n})^2}  + \frac{1}{2}
  f^2+ \dots \right]
\label{exp22}
\end{equation}
and the analogous ones for $\nu_f \rightarrow \nu_g$ together with the
following equation:
\begin{equation}
\lim_{\nu \rightarrow 0}  \frac{2 \sin \pi \nu}{\Theta_1 (  \nu |
  i t)} = - \frac{1}{ f_{1}^{3} (q)}~.
\label{the11}
\end{equation}
Notice that the factor in front of the bracket in
Eq. (\ref{closed}) can just be
computed for $f=g=0$,
because the corresponding quadratic term  in $f$ and $g$ has vanishing
coefficient as a consequence of the fact that the annulus diagram for
two undressed fractional branes vanishes.
One can then use
Eq. (\ref{f297}) for rewriting the combination $f^2 - g^2$ in terms of
the kinetic term of the gauge field and, after some calculation, one
obtains the following expression for the quadratic terms of the
gauge field:
\begin{eqnarray}
Z_h^c(F^2)\!\!\!&&=
\frac{N}{8 \pi^2} \int d^4 x \left( - \frac{1}{4} F_{\alpha \beta}^{a}
F^{ a \alpha  \beta }  \right)
\int_{0}^{\infty} \frac{dt}{t}  {e}^{- \frac{y^2}{2 \pi\alpha' t} }
\prod_{n=1}^{\infty}
\frac{ [1+ {q}^{ 2n-1 }  ]^4 [1+ {q}^{ 2n} ]^4 }{
[1- {q}^{2n-1}  ]^4 [1- {q}^{ 2n } ]^4} \times
\nonumber\\
&&\times \left[-1 + 8
\sum_{n=1}^{\infty}  \left( \frac{ {q}^{2n-1} }{
[1+ {q}^{2n-1}  ]^2}  - \frac{{q}^{2n}}{
[1+ {q}^{ 2n } ]^2}  \right)
 \right]
\nonumber\\
&&+\frac{iN}{32\pi^{2}}\int d^{4}x F_{\alpha\beta}^a\tilde F^{a\alpha\beta}
\int_0^\infty\frac{dt}{t}
e^{- \frac{y^2 }{ 2 \pi \alpha' t} }~.
\label{F289a}
\end{eqnarray}
By using the identities in Eq. (\ref{rel}), Eq. (\ref{F289a})
becomes:
\begin{eqnarray}
Z_h^c(F^2)\!\!\!&&= \frac{N}{8\pi^2 }
\int d^4 x \left( - \frac{1}{4} F_{\alpha \beta}^{a}
F^{ a \alpha  \beta }  \right)
\int_0^\infty
\frac{dt}{t}{\rm e}^{-\frac{y^2}{2\pi\alpha't}}
\left[\frac{f_3(q)f_2(q)}{f_1(q)f_4(q)}\right]^4
2q\frac{d}{dq} \log\left[\frac{f_3(q)}{f_2(q)}\right]\nonumber\\
&&
+\frac{iN}{32\pi^{2}}\int d^{4}x F_{\alpha\beta}^a\tilde F^{a\alpha\beta}
\int_0^\infty\frac{dt}{t}
e^{- \frac{y^2 }{ 2 \pi \alpha' t} }~~~,~~~ q ={e}^{- \pi t}
\label{g2}
\end{eqnarray}
that reproduces Eq. (\ref{F289}).
Using the modular transformations of the $f$-functions and the
relation
\begin{equation}
q \frac{d}{dq} = - \tau^2 k \frac{d}{dk}
\label{kqtau}
\end{equation}
one can
easily check that the previous equation is properly mapped in Eq.
(\ref{g1}) evaluated in the open channel by the open/closed string
duality.

\end{document}